\documentclass[aps,showpacs,onecolumn,superscriptaddress,12pt,nofootinbib]{revtex4-1}
\usepackage{epsfig,amssymb,amsmath,amsfonts,color}

\newcommand{\ket}[1]{\ensuremath{\left| #1 \right>}}

\newcommand{\average}[1]{\ensuremath{\left< #1 \right>}}

\begin{document}

\title{From thermal to excited-state quantum phase transitions ---the Dicke model}

\author{P. P\'erez-Fern\'andez}\email{pedropf@us.es}
\affiliation{Departamento de F\'{\i}sica Aplicada III, Escuela 
T\'ecnica Superior de Ingenier\'{\i}a, Universidad de Sevilla}

\author{A. Rela\~no}\email{armando.relano@gmail.com}
\affiliation{Departamento de F\'{\i}sica Aplicada I and GISC, Universidad Complutense de Madrid}
\date{today}

\begin{abstract}

  We study the thermodynamics of the full version of the Dicke model,
  including all the possible values of the total angular momentum $j$,
  with both microcanonical and canonical ensembles. We focus on both
  the excited-state quantum phase transition, appearing in the
  microcanonical description of the maximum angular momentum sector,
  $j=N/2$, and the thermal phase transition, which occurs when all the
  sectors are taken into account. We show that two different features
  characterize the full version of the Dicke model. If the system is
  in contact with a thermal bath and is described by means of the
  canonical ensemble, the parity symmetry becomes spontaneously broken
  at the critical temperature. In the microcanonical ensemble, and
  despite all the logarithmic singularities which characterize the
  excited-state quantum phase transition are ruled out when all the
  $j$-sectors are considered, there still exists a critical energy (or
  temperature) dividing the spectrum in two regions: one in which the
  parity symmetry can be broken, and another in which this symmetry is
  always well defined.

\end{abstract}

\maketitle

\section{Introduction}

Quantum phase transitions (QPTs) and critical phenomena play an
important role in the study of many-body quantum systems
\cite{Sachdev}. During the last decade, a new kind of phase transition
has been studied in depth ---the excited-state quantum phase
transition (ESQPT)
\cite{Caprio,Cejnar:06,Cejnar:08,Stransky:14,Stransky:15}. In
contrast to QPTs, which describe the non-analytical evolution of the
ground state energy as a function of a control parameter, ESQPTs
refer to a similar non-analytic behavior that takes place at a certain
critical energy $E_c$, when the control parameter responsible for the QPT is
kept fixed \cite{Heiss:02}. 

ESQPTs have been theoretically studied in many kinds of quantum
systems. Paradigmatic examples are the Lipkin-Meshkov-Glick (LMG)
\cite{LMG, LMG2}, the Dicke and Tavis Cummings models
\cite{Dicke,Brandes:13}, the interacting boson model \cite{IBM}, the
molecular vibron model \cite{Vibron}, atom-molecule condensates
\cite{Relano:14}, the kicked-top \cite{Bastidas:14} or the Rabi model
\cite{Richi:16}. Also, a number of experimental results have been
recently reported, in molecular systems \cite{Molecular},
superconducting microwave billiards \cite{Dietz:13}, and spinor
condensates \cite{Zhao:14}. However, and despite the intense research
performed during the last couple of years, some important questions
still remain open. The most important one is whether the critical
energy does separate two different phases in the spectrum. Contrary to
what happens in quantum and thermal phase transitions, there are no
clear traces of order parameters in ESQPTs. Though many physical
observables become singular at the critical point, it seems impossible
to find a magnitude which is zero at one side of the transition, and
remains different from zero at the other (see, for example
\cite{Stransky:14}). A recent proposal to characterize the transition
relies on how the values of the physical observables change with
energy \cite{Cejnar:16}. This idea allows us to identify two different
regions in the spectrum, but it does not provide an easy way to
distinguish different phases just by measuring an appropriate
observable.  Another recent proposal relies on symmetry-breaking. A
number of quantum systems showing ESQPTs are characterized by a
discrete $Z_2$ symmetry which can be broken at one side of the
transition, but not at the other. From fundamental physical reasons
this seems a promising idea. First, it links ESQPTs with the breakdown
of a certain symmetry, following a line of thought similar to the
theory of thermal phase transitions. Second, this fact entails
measurable dynamical consequences if a thermodynamic process is
performed from a symmetry-breaking initial condition ---the symmetry
of the final equilibrium state remains broken only if the final energy
is at the corresponding side of the transition, whereas the symmetry
is restored on the contrary \cite{Puebla:13, Puebla:15}. However, if
the initial condition has a well-defined value of this symmetry,
nothing similar happens when crossing the critical energy. In other
words, crossing an ESQPT does not entail a spontaneous breakdown of the
corresponding symmetry under any circumstances; the ocurrence of this
phenomenon depends on the details of the protocol.

Notwithstanding, the possible links between ESQPTs, thermal phase
transitions, and the breakdown of certain fundamental symmetries of the
system deserve to be explored. ESQPTs occur when the system is kept
isolated from any environment, and thus can be described by means of
the microcanonical ensemble. On the contrary, thermal phase
transitions take place at a certain critical temperature $\beta_c$,
and are usually described considering that the critical system is in
contact with a thermal bath, that is, by means of the canonical
ensemble \cite{Huang}. But, as microcanonical and canonical
descriptions become equivalent in the thermodynamical limit $N
\rightarrow \infty$, where $N$ is the number of particles of the
critical system, it is logical to expect that critical energy $E_c$
and critical temperature $\beta_c$ provide
  analogous information about the system. If we describe the critical
system by means of the canonical ensemble, we should expect that the
critical energy $E_c$ of the ESQPT corresponds to the internal energy
$U = - \partial \log Z / \partial \beta$, evaluated at the critical
temperature $\beta_c$, being $Z$ the canonical partition function. And
if the system is described by means of the microcanonical ensemble,
the critical temperature $\beta$ should correspond to the
microcanonical temperature $\beta = \partial \log \rho(E) / \partial
E$, evaluated at the critical energy $E_c$, being $\rho(E)$ the
density of states. However, all the facts discussed below suggest just
the opposite ---that thermal and excited-state quantum phase
transitions are totally different. Probably, this is due to the fact
that ESQPTs take place in systems with a small number of semiclassical
degrees of freedom, implying that the size of the corresponding
Hilbert space grows as $N^f$, being $f$ the number of degrees of
freedom ---the larger the number of degrees of freedom, the less
important are the consequences of the ESQPT \cite{Stransky:14}. On the
contrary, thermal phase transitions require an exponential growth of
the size of the Hilbert space with the number of particles, in order
to assure that intensive thermodynamical quantities, like the entropy
per particle $S/N$ or the Helmholtz potential per particle $F/N$, are
well defined in the thermodynamical limit. Hence, it is not clear even
whether the correspondence between thermal and excited-state quantum
phase transitions exists, or whether they are different phenomena
occurring under different physical circumstances. Indeed, it is shown in Ref. \cite{Stransky:17} that, for collective
  systems, the thermodynamical limit $N \rightarrow \infty$ does not
  coincide with the true thermodynamic limit unless the number of
  degree of freedom $f$ also tends to infinity. So, ESQPTs and thermal
  phase transitions appear for different asymptotic regimes of $N$ and
  $f$.  (We notice that, during the progress of this work, a similar
analysis, but with a different aim, was performed in the generalized
Dicke model, showing that it shows two different kinds of
superradiance \cite{Bastarrachea:16}).

In this work we tackle this task by studying, both analytically and
numerically, the Dicke model. It describes a system of $N$ two-level
atoms interacting with a single monochromatic electromagnetic
radiation mode within a cavity \cite{Dicke_original}. It is well known from the
seventies that this model exhibit a thermal phase transition
\cite{Carmichael:73, Comer-Duncan:74}. However, recently it was also
found that undergoes an ESQPT \cite{Dicke} aside the QPT \cite{Emery:03}. This kind of QPT
has been experimentally observed in several systems
\cite{Hartmann:06}, and the Dicke model itself can be simulated by
means of a Bose-Einstein condensate in an optical cavity
\cite{Baumann:10}. All these facts make this model the best one to
study the relationship between thermal and excited-state quantum phase
transitions. 

Up to now, the majority of the works on the Dicke model, including the
ones dealing with QPTs and ESQPTs (except \cite{Bastarrachea:16}, as
we have pointed above), were done in the subspace with maximum
pseudo-spin sector $j=N/2$, in which the ground state is
included. This restriction is enough to properly describe the recent
experimental results \cite{Baumann:10}, and also to study all the
consequences of the QPT. Furthermore, ESQPTs have been observed in the
subspace with $j=N/2$, which can be described by means of a
semiclassical approximation with just two degrees of freedom in the
thermodynamic limit $N \rightarrow \infty$. However, it is well known
that this restriction destroys the thermal phase transition
\cite{Aparicio:12}; the fact that the atomic subspace grows linearly
with $N$ in the $j=N/2$ sector makes impossible to properly define the
entropy $S$ or the Helmholtz potential $F$, and therefore precludes
the thermal phase transition. In this work we deal with the complete
Dicke model, including all the $j$ sectors. This is equivalent to increase the number of degree of freedom in the system that eventually goes to infinity in the thermodynamical limit.  Contrary to the seminal
papers on the thermal phase transition \cite{Carmichael:73,
  Comer-Duncan:74} we study the thermodynamics of this model in the
microcanonical ensemble, considering the system isolated instead of
being in contact with a thermal bath. This point of view allows us to study the possible connections between the excited-state and the thermal phase transitions. In particular, we
show that each $j$ sector displays the same kind of ESQPT, provided
that the coupling constant is large enough (see below for a detailed
discussion regarding this condition), but having each one a different
critical energy $E_c$.  Paradoxically, this fact, together with the
different weight of each $j$ sector in the spectrum of the complete
Dicke model, destroys most of the signatures of the ESQPT, and somehow
surprisingly entails the appearance of the typical signatures of
thermal phase transitions, like the existence of an order
parameter. In particular, we show that the collective contribution of
all the $j$ sectors rules out the logarithmic singularities in the
derivatives of the density of states, $\rho(E)$, and the third
component of the angular momentum, $J_z$, characteristic of the
ESQPT. However, one of the most important signatures of the ESQPT
survives. The parity symmetry of the Dicke model (see below for
details) can be still broken below the critical energy $E_c$,
which exactly coincides with the canonical internal
  energy, $U$, evaluated at the critical temperature of the thermal
  phase transition, $\beta_c$. In the microcanonical ensemble, that
  is, if the system remains isolated from any environment, the
expectation value of a symmetry-breaking observable, like $J_x$, is
always zero above $E_c$, but it can be different from zero
below; its particular behavior depends on the initial
  condition. If the system is in contact with a thermal bath, and is
  described by means of the canonical ensemble, a small
  symmetry-breaking term $\epsilon J_x $ produces that $\average{J_x}
  \neq 0$, even if we take the limit $\epsilon \rightarrow 0$ after
  the thermodynamical limit is done. In other words, the
  symmetry-breaking observable $J_x$ plays here the same role than the
  magnetization in the Ising model; it is an order parameter of the
  transition, and shows that the parity symmetry becomes
  spontaneoulsly broken below the critical temperature. On the other
  hand, the behavior of an isolated system is different. As it happens
  if only the highly-symmetric sector, $j=N/2$, is taken into account,
  the parity symmetry remains broken below the critical energy $E_c$
  of the ESQPT only if this symmetry is yet broken in the initial
  condition. In other words, the behavior of the system is expected to
  be different depending on whether the system is {\em heated} by
  means of the Joule effect, or it is in contact with a thermal
  bath.

This paper is organized as follows. In section $2$ we present the
Dicke model. In section $3$ we review the thermodynamics of the Dicke
model restricted to the highly-symmetric Dicke states, $\ket{j=N/2,
  M}$. We compare the results provided by the micro and canonical
ensembles, and we analyze the symmetry-breaking character of the
ESQPT. In section $4$ we perform a similar analysis including all the
$j$ sectors; we show that an ESQPT occurs in each $j$ sector. We also
show that the symmetry-breaking nature of the transition is still
present and leads to spontaneous symmetry-breaking if the system is in
contact with a thermal bath. In addition, we study the main physical
differences between the system in isolation and in contact with a
thermal bath. Finally, we extract the more relevant conclusions in the
last section.

\section{The Dicke model}

 The Dicke model describes the interaction of $N$
two-level atoms of splitting $\omega_0$ with a single bosonic mode of
frequency $\omega$, by means of a coupling parameter $\lambda$,
\begin{equation}
H = \omega_0 J_z + \omega a^{\dagger} a + \frac{2 \lambda}{\sqrt{N}} J_x \left( a^{\dagger} + a \right),
\label{eq:Dicke}
\end{equation}
where $a^{\dagger}$ and $a$ are the usual creation and annihilation
operators of photons, and $\vec{J} = (J_x ,J_y ,J_z )$ is the
Schwinger pseudospin representation of the $N$ two-level atom system,
that is, the total angular momentum of a system of $N$ $1/2$-spin
particles. This Hamiltonian has two conserved quantities. The first
one is $\Pi = \exp \left( i \pi \left[ j + J_z + a^{\dagger} a \right]
\right)$, due to the invariance of $H$ under $J_x \rightarrow - J_x$
and $a \rightarrow - a$; as this is a discrete symmetry, $\Pi$ has
only two different eigenvalues, $\Pi \left|E_i, \pm \right> = \pm
\left|E_i, \pm \right>$, and it is usually called {\em parity}. The
second one is the total angular momentum $J^2$ of the $N$ $1/2$-spin
particles. This entails that the Hamiltonian (\ref{eq:Dicke}) is
block-diagonal in $J^2$, and hence each sector is totally independent
from the others. The main dynamical consequence is that each $j$
sector evolves independently in any protocol keeping the Dicke model
isolated from any heat bath. Furthermore, as the recent experimental realizations
of this model involve only the sector of maximum angular momentum,
$j_{\text{max}}=N/2$, the great majority of the papers published
during the last couple of years are devoted to this case.

This model shows QPTs, ESQPTs and thermal phase transitions. In the
following paragraphs we summarize the known results.

\section{The case with $j=N/2$}

In this section, for the sake of completeness, we review the
thermodynamics of the Dicke model restricted to the highly-symmetric
Dicke states, $\ket{j=N/2, M}$. This configuration corresponds to a
two-level system in which $N$ bosons can occupy either the upper or
the lower level \cite{Aparicio:12}. It has been recently explored by
means of a Bose Einstein condensate in an optical cavity
\cite{Baumann:10}. First of all, we present the density of states
$\rho(E)$, which is computed by means of the microcanonical ensemble,
and later we show the same $\rho(E)$ but considering the calculation
in the canonical ensemble. These are well established
results. Finally, we compare both approaches and get some conclusions.

\subsection{Microcanonical ensemble}
\label{sec:jmax_micro}

Let's consider that the system is thermally isolated and that we
perform the following procedure: first, we freeze the system keeping
fixed all the external parameters of the Hamiltonian, up to it is
equilibrated at $T \sim 0$. This entails that the ground state, which
 always correspond with the sector of maximum angular momentum $j=N/2$, is the
only populated energy level. Second, we perform a quench, abruptly
changing one of the external parameters. Then, if the system remains
thermally isolated from the environment, the unitary evolution is
totally captured by the sector with $j=N/2$. Hence, all the
thermodynamic results after the system is equilibrated at the final
values of the external parameters should be obtained from a
microcanonical calculation with fixed $j=N/2$. This calculation can be
completed by means of a semiclassical approximation, following
different methods \cite{Altland:12, Brandes:13, Bastarrachea:14}. Here,
we follow the method in ref. \cite{Bastarrachea:14}.

Considering $\omega=\omega_0=1$, the density of states reads
\begin{equation}
\rho(E,j) =
\begin{cases}
2 j \; \text{ if } \; E/N > 1/2, \\
{\displaystyle \left( \frac{E}{j} + 1 \right) j + \frac{2}{\pi} \int_{E/j}^{y_+} \, dy \, \text{ acos} \left( \frac{ \sqrt{y - E/j}}{2 \lambda^2 (1-y^2)} \right) \text{ if } -1/2 \leq E/N \leq 1/2}, \\
{\displaystyle \frac{2 }{\pi} \int_{y_-}^{y_+} \, dy \, \text{ acos} \left( \frac{ \sqrt{y - E/j}}{2 \lambda^2 (1-y^2)} \right) \text{ if } E/N < - 1/2}, \\
\end{cases}
\label{eq:densidad}
\end{equation}
where 
\begin{equation}
y_- = - \frac{j + \sqrt{j} \sqrt{j + 8 E \lambda^2 + 16j \lambda^4}}{4 j \lambda^2},
\end{equation}
and
\begin{equation}
y_+ = \frac{-j + \sqrt{j} \sqrt{j + 8 E \lambda^2 + 16j \lambda^4}}{4 j \lambda^2}.
\end{equation}

For the third component of the angular momentum, we obtain
\begin{equation}
\frac{J_z(E,j)}{j} =
\begin{cases}
0 \; \text{ if } \; E/N > 1/2, \\
{\displaystyle \left( \frac{E^2}{2j^2} - \frac{1}{2} \right) \frac{j}{\rho(E,j)} + \frac{2j }{\pi \rho(E,j)} \int_{E/j}^{y_+} \, dy \, y  \text{ acos} \left( \frac{ \sqrt{y - E/j}}{2 \lambda^2 (1-y^2)} \right) \text{ if } -1/2 \leq E/N \leq 1/2}, \\
{\displaystyle \frac{2j}{\pi \rho(E,j)} \int_{y_-}^{y_+} \, dy \, y  \text{ acos} \left( \frac{ \sqrt{y - E/j}}{2 \lambda^2 (1-y^2)} \right) \text{ if } E/N < - 1/2}, \\
\end{cases}
\label{eq:jz}
\end{equation}

Finally, for the first component of the angular momentum and
considering that the parity is totally broken in the initial state,
\begin{equation}
\frac{J_x(E,j)}{j} =
\begin{cases}
0 \; \text{ if } \; E/N > -1/2, \\
{\displaystyle \pm \frac{2j}{\pi \rho(E,j)} \int_{y_-}^{y_+} \, dy \, \left(1 - y^2 \right)  \text{ acos} \left( \frac{ \sqrt{y - E/j}}{2 \lambda^2 (1-y^2)} \right) \text{ if } E/N < - 1/2}, \\
\end{cases}
\label{eq:jx}
\end{equation}
where the sign depends on the initial state. This expression has been
obtained taking into account only one of the two disjoint parts in
which the semiclassical phase space is divided for $\lambda>\lambda_c$
and $E<-N/2$ \cite{Puebla:13}. If the initial state has a
well-defined parity, both parts of the semiclassical phase space are
populated, giving rise to $\average{J_x}=0$.

These results show that an ESQPT happens at $E_c/N = - 1/2$
\cite{Dicke,Brandes:13,Puebla:13,Bastarrachea:14}. There
are singular points for both $\rho(E,j)$ and $J_z (E,j)$ ---the
derivatives of both magnitudes show a logarithmic divergence at
$E_c$. The reason for this behavior is the following: the density of
states, Eq. (\ref{eq:densidad}), is proportional to the size of the
phase space available to the system,
\begin{equation}
\rho(E,j) = C \int d q_1 d q_2 d p_1 d p_2 \, \delta \left[E - H(q_1, q_2; p_1, p_2) \right],
\end{equation}
where $q_1$ and $q_2$ denote the semiclassical coordinates; $p_1$ and
$p_2$, the semiclassical momenta, and $C$ is a normalization constant
(see, for example, \cite{Bastarrachea:14}). The key point is that despite this
semiclassical system is {\em finite}, it describes the quantum
Dicke model in the thermodynamical limit, $N \rightarrow \infty$, and it
has just $f=2$ degrees of freedom. Furthermore, every quantum system
showing an ESQPT is equivalent to a semiclassical system with a finite
number of degrees of freedom (see, for example, ref.
\cite{Stransky:14}). As a consequence, non-analyticities in the
quantum density of states are linked to stationary points in the
corresponding semiclassical model; and the geometric properties of
such stationary points determine the nature of the corresponding
singularities. In particular, systems with $f=1$ semiclassical degrees
of freedom show logarithmic singularities in the density of states, as well as
in certain physical observables at the critical energy of
the ESQPT, $E_c$; systems with $f=2$ degrees of freedom show
the same kind of singularities in the derivatives of the same
magnitudes \cite{Stransky:14}. Results for an higher number of degrees of freedom have been recently published, showing that the larger $f$, the higher the derivative in which the logarithmic singularity takes place \cite{Stransky:16}.

Also, if the parity symmetry is broken in the initial state,
$J_x(E,j)$ acts like an order parameter for the ESQPT; that is, it shows a
finite jump at $E_c$, from $\average{J_x} \neq 0$ to $\average{J_x} =
0$ \cite{Puebla:13}. On the contrary, initial conditions with well-defined positive (or negative) parity do not suffer any change when crossing the ESQPT. 

Another singular point is located at
$E_c/N=1/2$, whilst its critical character is controversial
\cite{Brandes:13,Bastarrachea:14}. Above this energy, $\rho(E)=1$ and
$\average{J_z}=0$, due to the ergodic character of the atomic
motion (now the whole phase space is accesible to the system). Despite this point is not usually identified as an ESQPT, it
has some of the features of a second order phase transition. First,
there exists an order parameter identifying two different phases: for
$E<N/2$, $\average{J_z} \neq 0$, whereas $\average{J_z}=0$ for
$E>N/2$. Second, there is a discontinuity in the derivative of
$\rho(E)$, that is, in the second derivative of the cumulated level
density $N(E)$. We will come back to this discussion in
Sec. \ref{sec:todo_j}. Numerical results illustrating these facts are
shown later.

\subsection{Canonical ensemble}

The same kind of calculation can be performed in the canonical
ensemble, considering that the system weakly interacts with a thermal
bath which commutes with $J^2$. Following ref. \cite{Aparicio:12} we can
obtain the partition function
\begin{equation}
Z(N,\beta)=\frac{1}{\pi} \int_{-\infty}^{\infty} d y \, \exp \left( - \beta \omega y^2 \right) \int_{-\infty}^{\infty} dx \, \exp \left( - \beta \omega x^2 \right) Z_g (N, \beta),
\end{equation}
where
\begin{equation}
Z_g (N, \beta) = \sum_{m=-N/2}^{N/2} \exp \left( - \beta m \sqrt{\omega_0^2 + \frac{4 \lambda^2 x^2}{N}} \right).
\end{equation}
The final result is
\begin{equation}
Z(N, \beta) = \sqrt{\frac{1}{\pi \beta \omega}} \int_{-\infty}^{\infty} d x \, \exp \left[ -\beta \left( \omega x^2 + \frac{n}{2} \sqrt{\omega_0^2 + \frac{4 \lambda^2 x^2}{n}} \right) \right] \frac{\exp \left( \beta (n+1)  \sqrt{\omega_0^2 + \frac{4 \lambda^2 x^2}{n}} \right) - 1}{\exp \left( \beta \sqrt{\omega_0^2 + \frac{4 \lambda^2 x^2}{n}} \right) -1}.
\end{equation}
There is no way to write this integral in terms of simple analytical
functions, but it can be evaluated numerically to obtain results for
precise values of all the external parameters $\omega$, $\omega_0$ and
$\lambda$. Furthermore, other thermodynamic results can be obtained
from the partition function
\begin{equation}
\average{E} = - \frac{\partial \log Z}{\partial \beta},
\label{eq:energia}
\end{equation}

\begin{equation}
\average{J_z}  =  - \frac{1}{\beta} \frac{\partial \log Z}{\partial \omega_0}.
\label{eq:jz2}
\end{equation}
In all the cases $\average{J_x} = 0$.

It has been shown that there is no thermal phase transition under
these circumstances \cite{Aparicio:12}. In other words, microcanonical
and canonical ensembles give rise to totally different results. If the
system remains thermally isolated there exists a critical energy $E_c
= -N/2$ at which a non-analiticity occurs, giving rise to a number of
dynamical (and observable) consequences \cite{Dicke,
  Puebla:13}. On the other hand, if the system is put in contact with
a thermal bath, everything changes smoothly with the temperature
$\beta$; in particular, nothing happens at the {\em critical}
temperature $\beta_c$, given by $\average{E(\beta_c)} = -N/2$.

\subsection{Results}

In this subsection, we compare the results of both the microcanonical and the
canonical calculations, for a system with $\omega=\omega_0=1$,
$\lambda = 3 \lambda_c = 1.5$, and $N=10^5$. All the results are
plotted versus the scaled energy $\average{E}/N$. For the canonical
calculation, this energy is obtained directly from
Eq. (\ref{eq:energia}).

\begin{figure}[h]
\includegraphics[scale=0.4,angle=-90]{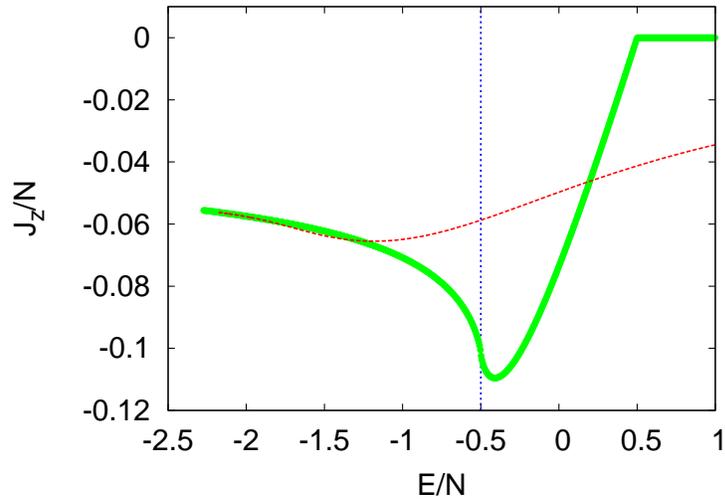}
\caption{(Color online) $J_z$ for the microcanonical ensemble (green solid points), and the canonical ensemble (dashed red line). The vertical dashed line
  shows the energy of the ESQPT.}
\label{fig:jz_jmax}
\end{figure}

\begin{figure}[h]
\includegraphics[scale=0.4,angle=-90]{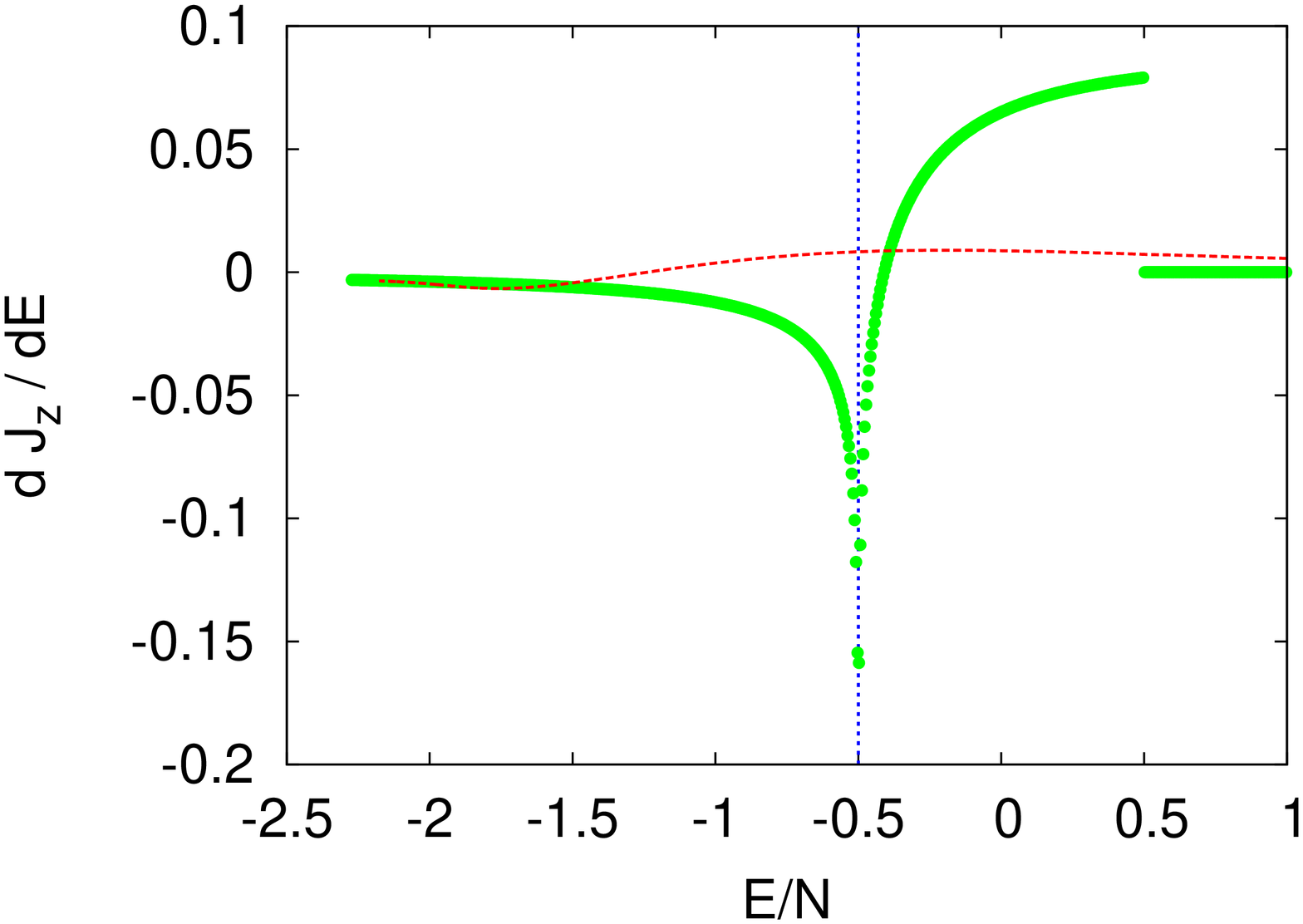}
\caption{(Color online) $d J_z/ d E$ for the microcanonical ensemble (green solid points), and
  the canonical ensemble (dashed red line). The vertical dashed line
  shows the energy of the ESQPT.}
\label{fig:derjz_jmax}
\end{figure}

\begin{figure}[h]
\includegraphics[scale=0.4,angle=-90]{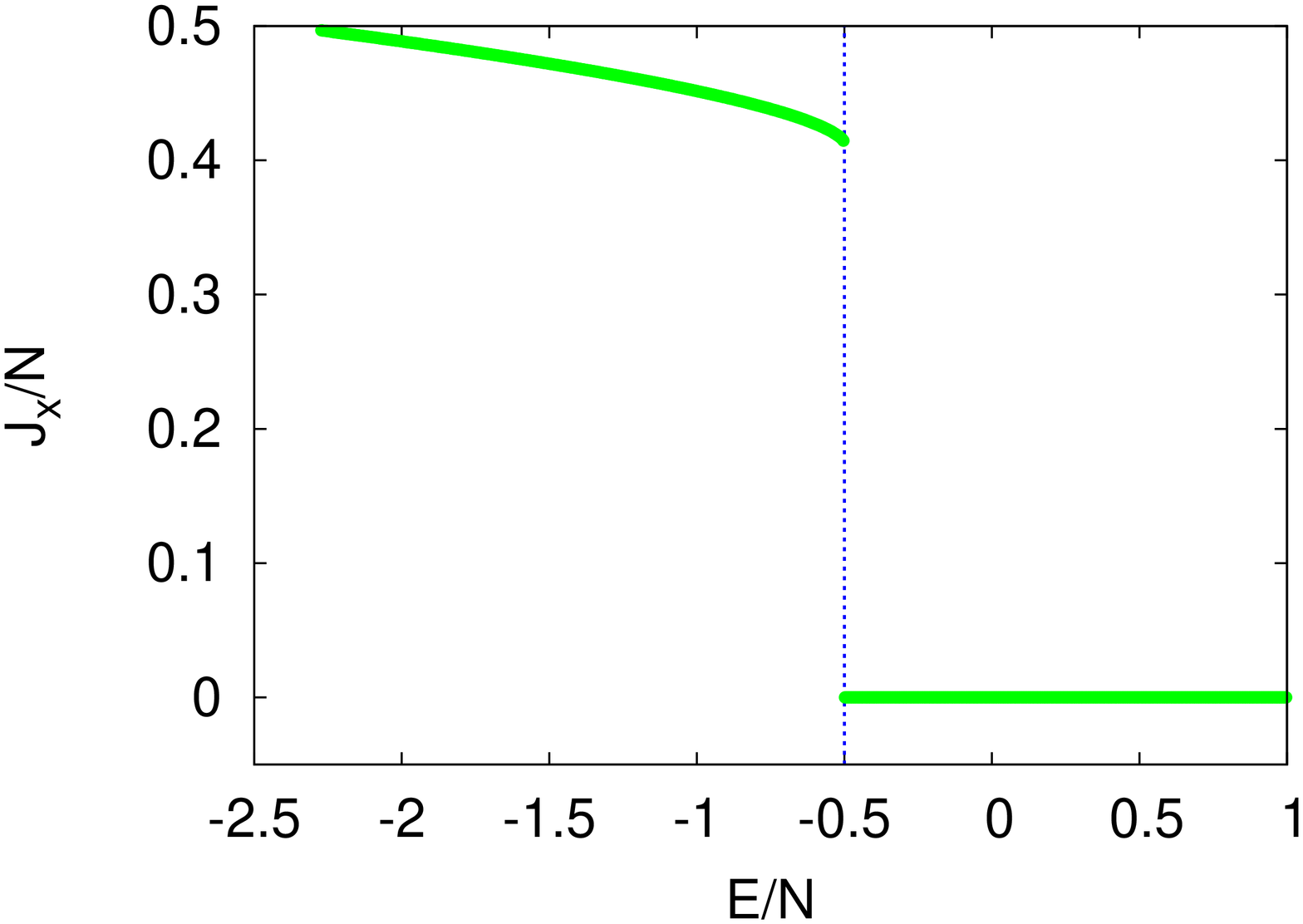}
\caption{(Color online) $J_x$ for the microcanonical ensemble (green solid points). The vertical dashed line
  shows the energy of the ESQPT.}
\label{fig:jx_jmax}
\end{figure}

In Figs. \ref{fig:jz_jmax}, \ref{fig:derjz_jmax}, and
\ref{fig:jx_jmax} we depict the results for $\average{J_z}$, $\average{d
  J_z / d E}$, and $\average{J_x}/N$ respectively. In the first two cases, we show
both the microcanonical (solid green points) and the canonical (dashed
red line) calculations; in Fig. \ref{fig:jx_jmax}, we show just the
microcanonical calculation, because $\average{J_x}=0$ in the canonical
ensemble. In all the cases we show the critical energy of the ESQPT,
$E_c/N=-1/2$, by means of a vertical dashed line.

As a general result, we can observe that the behavior of the Dicke
model in the $j=N/2$ sector is totally different depending whether it
is thermally isolated or in contact with a thermal bath. In the first
case, we can see neat signatures of the ESQPT (a singular point in
$\left< d J_z /d E (E_c) \right>$, or the crossing from $\left< J_x(E)
\right> \neq 0$ to $\left< J_x(E) \right> = 0$, if the parity symmetry
is broken in the initial state). In the second one, no traces of such
phenomena are present. The reason behind this result is that the
microcanonical density of states grows linearly with $N$. This entails
that the thermodynamic magnitudes that should be extensive, like the
entropy, $S$, or the Helmholtz free energy, $F$, grow with $\log N$,
and therefore $S/N \rightarrow 0$ and $F/N \rightarrow 0$ in the
thermodynamic limit. The main consequence is that the different
ensembles are not equivalent in the thermodynamic limit, and that
thermodynamics in this system is far from usual, and hence the results
for the different statistical ensembles do not coincide.  As it is
pointed in \cite{Stransky:17} this is due to the finite number of
(semiclassical) degrees of freedom that the system has in the
thermodynamic limit, $N \rightarrow \infty$.

\section{The full Dicke model}
\label{sec:todo_j}

In this section, we perform a similar analysis as the former one, but
now including all the $j$-sectors in the calculation. From a
semiclassical point of view, this entails that the number of degrees
of freedom $f$ also goes to infinity in the thermodynamical limit.

\subsection{Microcanonical ensemble}

If we consider that the system is thermally isolated, we can follow
the same procedure than for the case with $j=N/2$, taking into account
that each $j$-sector is totally independent from the others. In other
words, we can rely on the semiclassical approximation for each $j$
sector, and then collect all these results. Note, however, that the
semiclassical approximation only gives good results for large values
of the total number of two-level atoms, $N$. Hence, our procedure is
questionable for sectors with low values of $j$, and, in
particular, for the $j=0$ sector. This issue is discussed in detail
later on.

To profit from the results obtained in the previous section, we proceed in the following way. The full Dicke model reads,
\begin{equation}
H = \omega a^{\dagger} a + \omega_0 J_z + \frac{2 \lambda}{\sqrt{N}} J_x \left( a + a^{\dagger} \right).
\end{equation}
Considering that this Hamiltonian is block-diagonal in a $\left|j, M \right>$ basis, the previous equation can be written as follows,
\begin{equation}
H_j = \omega a^{\dagger} a + \omega_0 J_z + \frac{2 \lambda}{\sqrt{2 j}} \sqrt{\frac{2j}{N}} J_x \left( a + a^{\dagger} \right),
\end{equation}
where $H_j$ denotes the Hamiltonian $H$ in the sector with total
angular moment equal to $j$. Thus, we can define an effective coupling
constant for each $j$-sector, $\lambda^{\text{eff}}_j = \lambda
\sqrt{2j/N}$, giving rise to
\begin{equation}
H_j = \omega a^{\dagger} a + \omega_0 J_z + \frac{2 \lambda^{\text{eff}}_j}{\sqrt{2 j}} J_x \left( a + a^{\dagger} \right).
\end{equation}
From this result we conclude that the Hamiltonian of each $j$-sector, $H_j$, is formally identical than the one of the highly-symmetric sector, $j=N/2$, but with a different effective coupling $\lambda^{\text{eff}}_j$.

With this in mind, we proceed to discuss the presence of ESQPTs in
each $j$-sector. From the results derived in the section
\ref{sec:jmax_micro}, we conclude:

\begin{enumerate}

\item ESQPT appears if $\lambda > \lambda_c = \sqrt{\omega
    \omega_0}/2$. This entails that each $j$-sector requires a different coupling constant to show the ESQPT, the smaller the value of $j$, the larger the coupling,
\begin{equation}
\lambda_c^{(j)} = \sqrt{\frac{N \omega \omega_0}{8 j}}.
\end{equation}
Therefore, the $j=0$ sector does not exhibit an ESQPT in any case
($\lambda_c^{(j)} \rightarrow \infty$), and the lower values of $j$
require so large coupling constants for having ESQPTs, that these
transitions are restricted to the larger values of $j$ in all the
practical cases.

\item The critical energy for each sector is located at $E_c^j/N = -j
  /N$, and the energy of the other singular point at
  $E_*^j/N=j/N$. Thus, the lower $j$, the smaller is the energy band
  between these two singular points. If $j \rightarrow 0$ with a
  coupling constant large enough for the ESQPT to occur, the band
  shrinks to a single point located at $E/N=0$.

\item For any finite value of the coupling strenght in the
  superradiant phase, $\lambda > \lambda_c$, the dynamics of the full
  Hamiltonian is the result of collecting all the $j$-sectors, with
  both critical and non-critical behaviour.

\end{enumerate}

Considering that each $j$-sector is totally independent from the
others, the density of states for the full Hamiltonian can be obtained as
\begin{equation}
\rho(E) = \sum_{j=0}^{N/2} g(N,j) \rho(E,j),
\end{equation}
being $g(N,j)$ the degeneracy of each $j$-sector, and $\rho(E,j)$ is
given by Eq. (\ref{eq:densidad}).

The degeneracy is obtained as the number of ways in which a set of $N$
$1/2$-spin particles can give rise to a total angular momentum
$j$. The result is
\begin{equation}
g (N,j) = \frac{1 + 2j}{1 + j + N/2} \begin{pmatrix} N \\ N/2 -j \end{pmatrix}.
\end{equation}
To make easier the analytical calculations, it is preferable to work with an alternative version of this expresion. Instead of the angular momentum $j$, we consider the variable $x=j/N$, which can be taken as a continuous variable $x \in [0,1/2]$ in the thermodynamical limit, $N \rightarrow \infty$. Also, we write the combinatorial numbers in terms of the Gamma function, and therefore we obtain a continuous function $g(N,x)$ for any finite (but large) value of $N$,
\begin{equation}
g(N,x) = \frac{\left(1 + 2 N x \right) \Gamma(N+1)}{\Gamma \left( 1 + N/2 - Nx \right) \Gamma \left( 2 + N/2 + Nx \right)}.
\label{eq:deg}
\end{equation}
Hence, the total density of states is given by
\begin{equation}
\rho(E,N) = \int_0^{1/2} d x \, g(N,x) \rho(E,Nx).
\end{equation}

We can apply the same procedure to the expected values of $J_z$ and
$J_x$, obtaining
\begin{equation}
J_z (E,N) = \frac{1}{\rho(E,N)} \int_0^{1/2} dx \, g(N,x) \rho(E,Nx) J_z(E, Nx),
\label{eq:micro_jz}
\end{equation}
with $J_z(E,Nx)$ given by Eq. (\ref{eq:jz}). And
\begin{equation}
J_x (E,N) = \frac{1}{\rho(E,N)} \int_0^{1/2} dx \, g(N,x) \rho(E,Nx) J_x(E, Nx),
\label{eq:micro_jx}
\end{equation}
with $J_x(E,Nx)$ given by Eq. (\ref{eq:jx}). All these integrals have
to be performed numerically since it is not possible to get analytical expressions.

As it has been pointed before, this procedure assumes that all the $j$
sectors can be properly described by means of the semiclassical
approximation, and this is not completely true. Therefore, the
goodness of the final result critically depends on the shape of
Eq. (\ref{eq:deg}). If the subsequent integrals are dominated by
sectors with $j$ large enough, we can rely on our procedure; if they
are dominated by the lowest $j$-sectors, the procedure is not going to
work. So, prior to present the numerical results, we study here the
shape of the function $g(N,x)$. Its maximum can be obtained by solving the equation $d g(N,x) / dx =0$. An asymptotic expansion when $N \rightarrow \infty$ and a Taylor expansion around $x =0$ show that this maximum is located at
\begin{equation}
x_{\text{max}} = \frac{1}{2 \sqrt{N}} - \frac{1}{2N} + O \left( \frac{1}{N^{3/2}} \right).
\end{equation}
This result imply two apparently contradictory consequences. First, the maximaly degenerated $j$-sector is
\begin{equation}
j_{\text{max}} = \frac{\sqrt{N}}{2} - \frac{1}{2} + O \left( \frac{1}{\sqrt{N}} \right). 
\end{equation}
Hence, $j_{\text{max}} \rightarrow \infty$ in the thermodynamical
limit, and consequently the semiclassical approximation used to derive
Eqs. (\ref{eq:densidad}), (\ref{eq:jz}) and (\ref{eq:jx}) is expected
to work provided that $N$ is large enough. On the contrary, it is also
true that $x_{\text{max}} \rightarrow 0$ when $N \rightarrow \infty$,
suggesting that the maximally degenerated sector, and the one
responsible of the behavior of the full Dicke Hamiltonian, is $j
\rightarrow 0$, or the corresponding to the lower value of $j$
compatible with the given energy \cite{Bastarrachea:16}. The solution
of this apparent paradox is that $g(N,x)$ becomes non-continuous in
the thermodynamical limit. Exact calculations from Eq. (\ref{eq:deg})
show that
\begin{eqnarray}
g(N,0) &=& \frac{2 \Gamma \left(1+N\right)}{\left(2+N \right) \Gamma \left(1 + N/2 \right)^2}, \\
g(N,x_{\text{max}}) &=& \frac{N^{3/2} \Gamma \left(N\right)}{\Gamma \left(\frac{3 -\sqrt{N} + N}{2} \right) \Gamma \left(\frac{3 +\sqrt{N} + N}{2} \right) }. 
\end{eqnarray}
And the corresponding asymptotic expansion when $N \rightarrow \infty$
give rise to
\begin{eqnarray}
g(N,0) &\approx& \frac{2^{3/2}}{\sqrt{\pi}} \frac{2^N}{N^{3/2}}, \\
g(N,x_{\text{max}}) &\approx& \frac{2^{3/2}}{\sqrt{\pi}} \frac{\text{e}^{-1/2}2^N}{N}. 
\end{eqnarray}
Therefore, the degeneracy of the $j_{\text{max}}$-sector is larger
than the degeneracy of the sector with $j=0$ for any finite size
system with $N$ atoms, and the corresponding ratio is
\begin{equation}
\frac{g(N,0)}{g(N,x_{\text{max}})} \approx \frac{\text{e}^{1/2}}{\sqrt{N}} \rightarrow 0, \; \; \text{ when } N \rightarrow \infty.
\end{equation}
In other words, $\lim_{N \rightarrow \infty} g(N, x_{\text{max}}) \neq g(N,0)$ despite $\lim_{N \rightarrow \infty} x_{\text{max}} = 0$, implying that $g(N,x)$ becomes non-continuous in the thermodynamical limit. Therefore, a rigurous calculation of the full density of states $\rho(E,N)$ and the corresponding expected values $J_z(E,N)$ and $J_x (E,N)$, requires to take this fact into account. Notwithstanding, from a practical point of view this is only important if we are interested in finite-size systems, or in obtaining finite-size corrections to the behavior in the thermodynamical limit. A first-order approximation for the behavior in the thermodynamical limit can be obtained just by considering the lower $j$-sector existing at a given energy $E$, which coincides with $j=0$ for $E/N>0$ \cite{Bastarrachea:16}. In the next sections we will provide numerical results illustrating all these facts.

\subsection{Canonical ensemble}

Let's consider that the system is in contact with a thermal bath, so
the total Hamiltonian (system $+$ environment) reads
\begin{equation}
H = H_{\text{Dicke}} + H_{\text{bath}} + H_I,
\end{equation}
where $H_I$ is the interacting term between the system (the Dicke
model) and its environment. If we assume that $[H_I,J^2] \neq 0$ and
$[H_I, \Pi] \neq 0$, we have to take into account both parities and
all the possible values of the angular momentum to derive the
thermodynamics of the Dicke model. As it is indicated in
\cite{Aparicio:12}, this is equivalent to a set of $N$ fermions
occupying either the lower or the upper level of a two-level
system. Under such circumstances, the partition function can be
explicitely obtained; this calculation was completed around $40$ years
ago \cite{Comer-Duncan:74}. Here, we summarize the main results.

The partition function can be exactly derived, giving rise to
\begin{equation}
Z(N,\beta) = \frac{2^N}{\sqrt{\pi \beta \omega}} \int_{-\infty}^{\infty} \, dx \exp \left( -\beta \omega x^2 \right) \left[ \text{cosh} \left( \frac{\beta \sqrt{N \omega_0^2 + 16 \lambda^2 x^2}}{2 \sqrt{N}} \right)\right]^N.
\end{equation}
This integral cannot be solved in terms of simple analytical
functions. Exact results have to be derived by means of numerical
integration. The same procedure can be used to obtain the expected
values of the relevant observables of the system. For example, we can
obtain $J_x$ and $J_z$ considering
\begin{equation}
J_{\alpha} (N,\beta) = \frac{1}{Z(N,\beta)}\text{Tr} \left[ J_{\alpha} \exp \left(-\beta H \right) \right],
\end{equation}
\noindent where $\alpha= x, y, z$ is a label.
From this equation it is straightforward to obtain
\begin{equation}
J_z(N, \beta) = - \frac{\omega_0 2^{N-1}}{Z (N,\beta)}  \sqrt{\frac{N^3}{\pi \beta \omega}} \int_{-\infty}^{\infty} d x \, \exp \left( - \beta \omega x^2 \right) \frac{\left[ \text{cosh} \left( \frac{\beta \sqrt{N \omega_0^2 + 16 \lambda^2 x^2}}{2 \sqrt{N}} \right)\right]^{N-1} \text{sinh} \left( \frac{\beta \sqrt{N \omega_0^2 + 16 \lambda^2 x^2}}{2 \sqrt{N}} \right)}{\sqrt{N \omega_0^2 + 16 \lambda^2 x^2}}
\end{equation}
\begin{equation}
J_x(N, \beta) = - \frac{N \lambda 2^{N-1}}{Z (N,\beta)}  \sqrt{\frac{1}{\pi \beta \omega}} \int_{-\infty}^{\infty} d x \, x \exp \left( - \beta \omega x^2 \right) \frac{\left[ \text{cosh} \left( \frac{\beta \sqrt{N \omega_0^2 + 16 \lambda^2 x^2}}{2 \sqrt{N}} \right)\right]^{N-1} \text{sinh} \left( \frac{\beta \sqrt{N \omega_0^2 + 16 \lambda^2 x^2}}{2 \sqrt{N}} \right)}{\sqrt{N \omega_0^2 + 16 \lambda^2 x^2}}
\end{equation}
Note that the last integral is an odd function in the $x$ variable, so $J_x (N,
\beta) =0$. The same happens for any other symmetry-breaking
observable, like, for example $q=(a+a^{\dagger})/2$. Also, both
$\average{E}$ and $\average{J_z}$ can be obtained directly from the
partition function making use of Eqs. (\ref{eq:energia}) and
(\ref{eq:jz2}).

Since phase transitions are defined in the thermodynamic limit, $N
\rightarrow \infty$, we can apply Laplace's method to evaluate the
partition function. Defining $y^2 = x^2/N$ we can write
\begin{equation}
Z(N,\beta) = \frac{\sqrt{N}}{\sqrt{\pi \beta \omega}} \int_{-\infty}^{\infty} \, dy \exp \left\lbrace N \left[ -\beta \omega y^2 + \log \left( 2 \text{cosh} \left[ \frac{\beta \omega_0}{2} \sqrt{1 + \frac{16 \lambda^2 y^2}{\omega_0^2}} \right] \right) \right] \right\rbrace.
\end{equation}
As a consequence,
\begin{equation}
\lim_{N \rightarrow \infty} Z(N,\beta) = \sqrt{\frac{2}{\beta \left| \Psi''(y_0) \right|}} \exp \left[ N \Psi(y_0) \right],
\end{equation}
where
\begin{equation}
\Psi(y)= -\beta \omega y^2 + \log \left( 2 \text{cosh} \left[ \frac{\beta \omega_0}{2} \sqrt{1 + \frac{16 \lambda^2 y^2}{\omega_0^2}} \right] \right),
\end{equation}
and $y_0$ is the value of $y$ which maximizes $\Psi(y)$. 

A phase transition normally happens when the possition of the maximum
$y_0$ changes at a certain critical tempertaure $\beta_c$. The easiest
way to obtain $y_0$ is solving $\Psi'(y_0)=0$, and evaluating
$\Psi(y_0)$ for all the solutions. For the Dicke model, the trivial
solution $y_0=0$ exists for all the temperatures and the values of the system
parameters. Under certain circumstances, there also exists another solution,
\begin{equation}
  \frac{4 \lambda^2}{\omega} \text{tanh} \left( \frac{\beta \omega_0}{2} \sqrt{1 + \frac{16 \lambda^2 y_0^2}{\omega_0^2}} \right) = \omega_0 \sqrt{1 + \frac{16 \lambda^2 y_0^2}{\omega_0^2}}.
\end{equation}
Defining $z=\sqrt{1 + 16 \lambda^2 y_0^2 / w_0^2}$, the
former equation reads,
\begin{equation}
\text{tanh} \left( \frac{\beta \omega_0 z}{2} \right) = \frac{\omega \omega_0}{4 \lambda^2} z.
\label{eq:critico}
\end{equation}
It is important to note that, by definition, $z>1$.

As $ -1 < \text{tanh} (z) < 1$ $\forall z$, the former equation only has solutions if 
\begin{equation}
\lambda > \lambda_c = \frac{\sqrt{\omega \omega_0}}{2}.
\end{equation}
Furthermore, the only way for Eq. (\ref{eq:critico}) having a solution
for $z>1$ is that $\text{tanh} \left( \frac{\beta z}{2} \right) >
\frac{\omega \omega_0}{4 \lambda^2} z$ at $z=1$; if this condition
does not hold, the right side of the equation is larger than the left
for any $z>1$. Therefore, if
\begin{equation}
\beta < \frac{2}{\omega_0} \, \text{tanh}^{-1} \left( \frac{\omega \omega_0}{4 \lambda^2} \right),
\end{equation}
the only solution of the problem is the trivial one $y_0 = 0$. On the
contrary, if $\beta$ exceeds this value, there exists a non-trivial
solution $\widetilde{y}_0 \neq 0$. Evaluating $\Psi(0)$ and
$\Psi(\widetilde{y}_0)$ we can see that $\Psi(\widetilde{y}_0) >
\Psi(0)$ in all the cases. Therefore, the position of the maximum
$y_0$ changes at the critical temperature
\begin{equation}
\beta_c = \frac{2}{\omega_0} \, \text{tanh}^{-1} \left( \frac{\omega \omega_0}{4 \lambda^2} \right),
\label{eq:critica}
\end{equation}
entailing that the partition function becomes non-analytic at the
critical temperature $\beta_c$.

Summarizing, if $\lambda < \lambda_c$, there is no thermal phase
transition. At $\lambda = \lambda_c$, the phase transition takes place at
$\beta \rightarrow \infty$, that is, at $T \rightarrow 0$; it
constitutes a QPT. If $\lambda>\lambda_c$, there exists a thermal
phase transition at a critical temperature $T_c = 1/ \beta_c$. The
values for $\average{E}$ and $\average{J_z}$ in the thermodynamical
limit can be easily obtained making use of Eqs. (\ref{eq:energia}) and
(\ref{eq:jz2}).

\subsection{Spontaneous symmetry-breaking at the critical temperature}
\label{sec:symmetry_breaking}

Phase transitions are usually linked to the breakdown of a global
symmetry of the Hamiltonian. Above the critical temperature, the
stable phase have the same symmetries than the Hamiltonian; below, one
of these symmetries becomes spontaneously broken. The main signature
of this fact usually lays in the behavior of the order parameter. For
example, the paradigmatic Ising model without external magnetic field
is symmetric under the permutation of all the spins, but the system
becomes spontaneously magnetizated below the critical temperature. The
usual order parameter of this transition reflects this fact. In any
symmetric state, the total magnetization $m=M/N$ is zero; however, $m$
becomes different from zero in the ferromagnetic phase.

The seminal papers on the superradiant phase transition in the Dicke
model do not consider this feature. As we have discussed above, the
Dicke model has a discrete $Z_2$ symmetry, the parity $\exp\left( i
  \pi \left[ j + J_z + a^{\dagger} a \right] \right)$. The usual
order parameters for the superradiant transition are either $J_z$ or
$a^{\dagger} a$. These observables provide a good physical insight of
the character of the transition: in the superradiant phase both the
bosonic field and the upper level of the atomic system are macroscopically
populated, even when $\beta \rightarrow \infty$, given rise to expected values
$\left<J_z \right>$ and $\left< a^{\dagger} a \right>$ different from zero
\cite{Carmichael:73,Comer-Duncan:74,Emery:03}. However, neither $J_z$
and $a^{\dagger} a$ break the parity symmetry. Thus, it is interesting
to seek alternative order parameters playing the same role than the
magnetization in the Ising model. A good one is $J_x$, which has been
recently used to study the ESQPT in the highly-symmetric sector
\cite{Puebla:13}. As $\left< J_x \right>=0$ in any eigenstate with
well-defined parity, the strategy to study the behavior of this
observable when crossing the phase transition consists in introducing
a small symmetry breaking term in the Hamiltonian,
\begin{equation}
H_{\epsilon} = \omega a^{\dagger} a + \omega_0 J_z + \frac{2 \lambda}{\sqrt{N}} J_x \left( a^{\dagger} + a \right) + \epsilon J_x,
\end{equation}
and taking $\epsilon \ll \omega, \omega_0, \lambda$.

The partition function of this system can be obtained following the
same strategy than in the previous section. In the thermodynamical
limit,
\begin{equation}
\lim_{N \rightarrow \infty} Z_{\epsilon} (N,\beta) = \sqrt{\frac{2}{\beta \left| \Psi_{\epsilon}''(y_0) \right|}} \exp \left[ N \Psi_{\epsilon} (y_0) \right],
\end{equation}
where
\begin{equation}
\Psi_{\epsilon}(y) = - \beta \omega y^2 + \log \left( 2 \cosh \left[ \frac{\beta \omega_0}{2} \sqrt{1 + \left( \frac{\epsilon + 4 \lambda y}{\omega_0} \right)^2} \right] \right),
\end{equation}
and $y_0$ is the value that maximizes $\Psi_{\epsilon}(y)$. From this result, we can obtain the expected values of $J_z$ and $J_x$ by means of
\begin{eqnarray}
\left< J_z \right>_{\epsilon} &=&  \frac{1}{\beta} \frac{\partial \log Z_{\epsilon}}{\partial \omega_0}, \\
\left< J_x \right>_{\epsilon} &=&  \frac{1}{\beta} \frac{\partial \log Z_{\epsilon}}{\partial \epsilon}.
\end{eqnarray} 
And finally, we can study both parameters in the limit
$\epsilon \rightarrow 0$. It is worth to remark that this procedure entails that
the thermodynamic limit is taken {\em before} the $\epsilon
\rightarrow 0$ limit. Spontaneous symmetry breaking in phase
transitions occurs because these limits do not commute, leading to a
finite value of the symmetry-breaking order parameter even in the
limit $\epsilon \rightarrow 0$.

\begin{figure}[h!]
\includegraphics[scale=0.3,angle=-90]{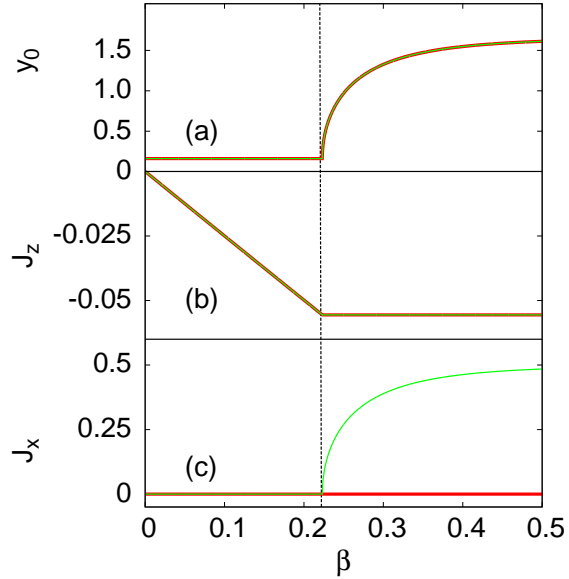}
\caption{(Color online) Value of $y_0$ and expected values of $J_z$ and $J_x$, both without the symmetry-breaking term (thick line, red online) and with the symmetry-breaking term, taking the limit $\epsilon \rightarrow 0$ (thin line, green online). Critical temperature $\beta_c$ is marked with a dotted vertical line.}
\label{fig:symetry1} 
\end{figure}

In Fig. \ref{fig:symetry1} we show the results for $y_0$, $\left< J_z
\right>$ and $\left< J_x \right>$, both for the normal Dicke model,
and for the case with the symmetry-breaking term, considering the
limit $\epsilon \rightarrow 0$ (see caption for details). We can see
that including the symmetry-breaking term does not change the results
for the critical temperature, $\beta_c$, the value of $y_0$ and the
expected value for $J_z$. However, $\left< J_x \right>$ changes
dramatically: it is identically zero at both sides of the transition
if the symmetry-breaking term is not included, but becomes different
from zero in the superradiant phase if it is included, even if we take
the $\epsilon \rightarrow 0$ limit. Hence, we conclude that the parity
symmetry is spontaneously broken for $\beta > \beta_c$, and that $J_x$
is a good order parameter of the transition. Furthermore, this
observable plays the same role than the magnetization in the
paradigmatic Ising model.

Summarizing, from the results shown in this section we conclude that
$J_x$ is the proper order parameter for the superradiant phase
transition. In the following sections, we will compare this finding
and the recently published results about symmetry-breaking and the
ESQPT \cite{Puebla:13}.

\subsection{Numerical results: different $j$-sectors}

\begin{figure}[h!]
\begin{tabular}{cc}
\includegraphics[scale=0.3,angle=-90]{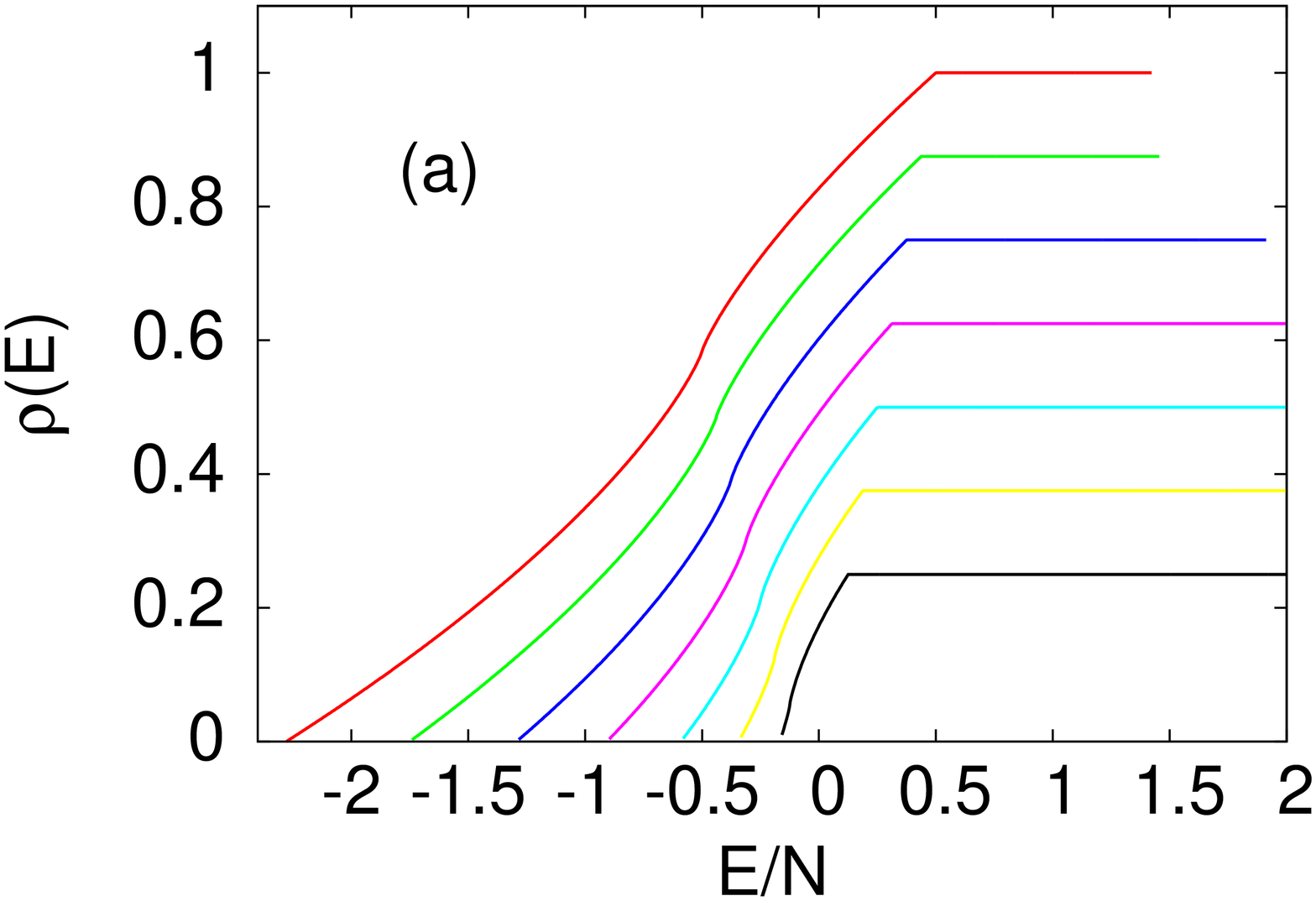} &
\includegraphics[scale=0.3,angle=-90]{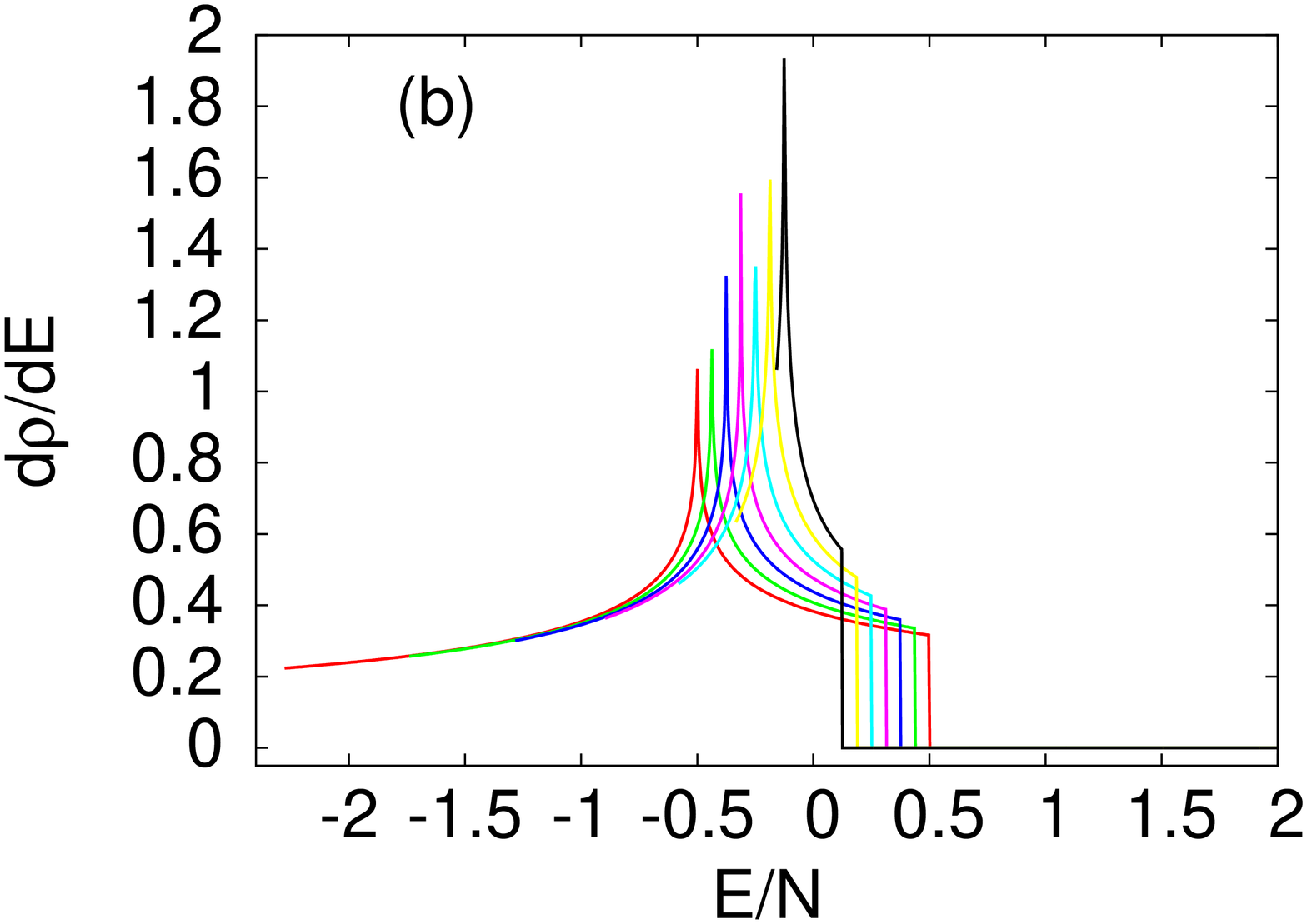} \\
\includegraphics[scale=0.3,angle=-90]{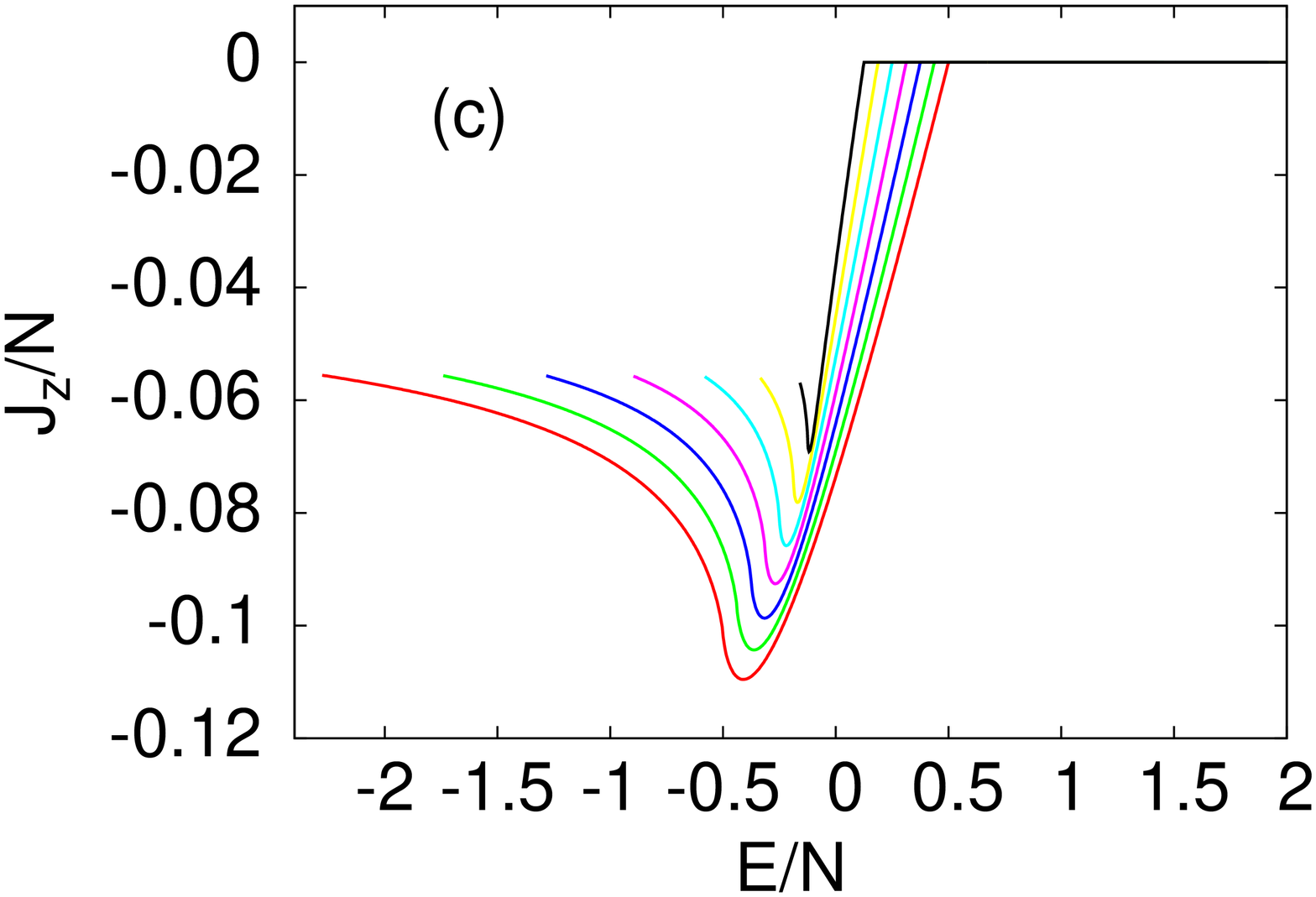} &
\includegraphics[scale=0.3,angle=-90]{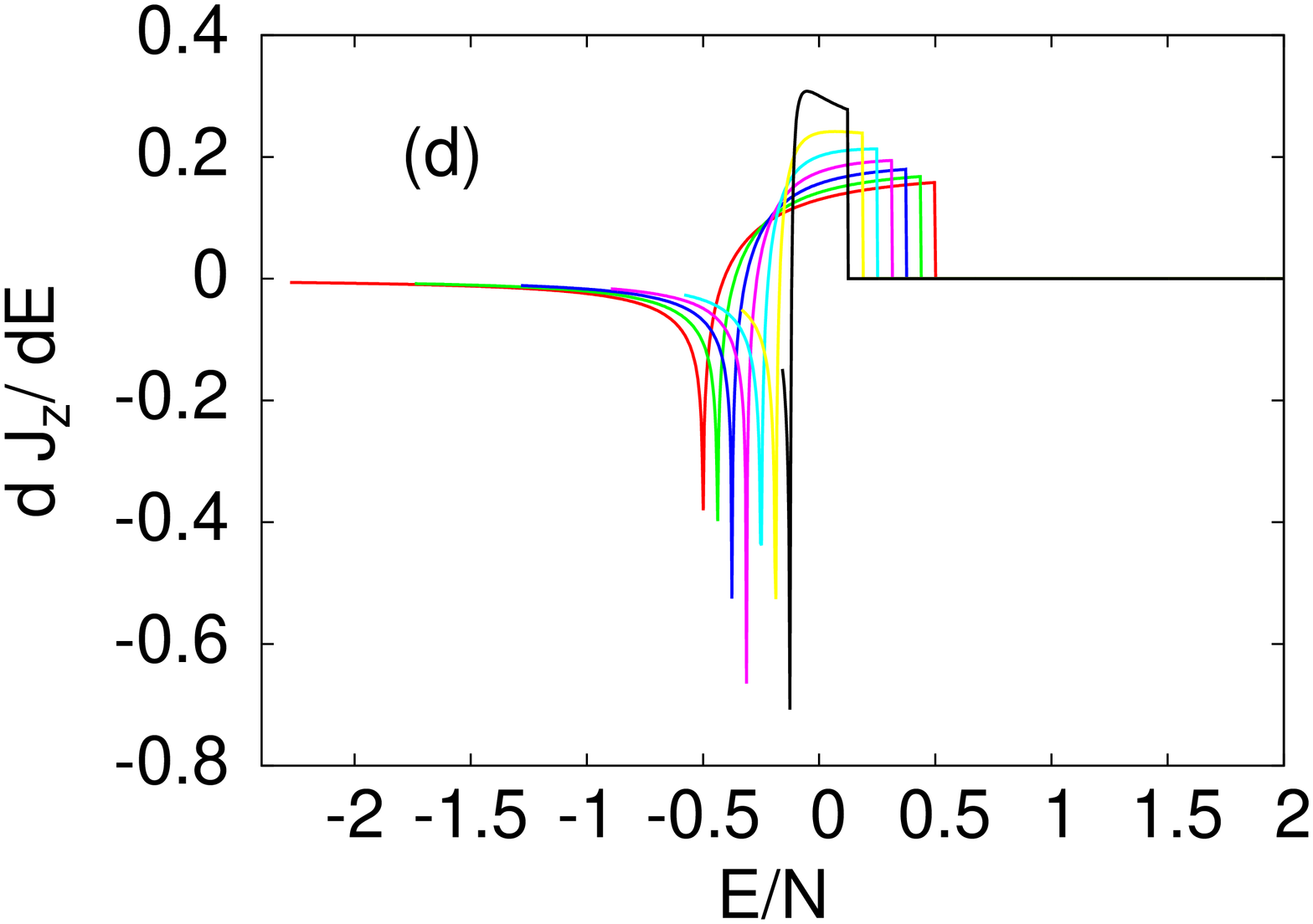} \\
\includegraphics[scale=0.3,angle=-90]{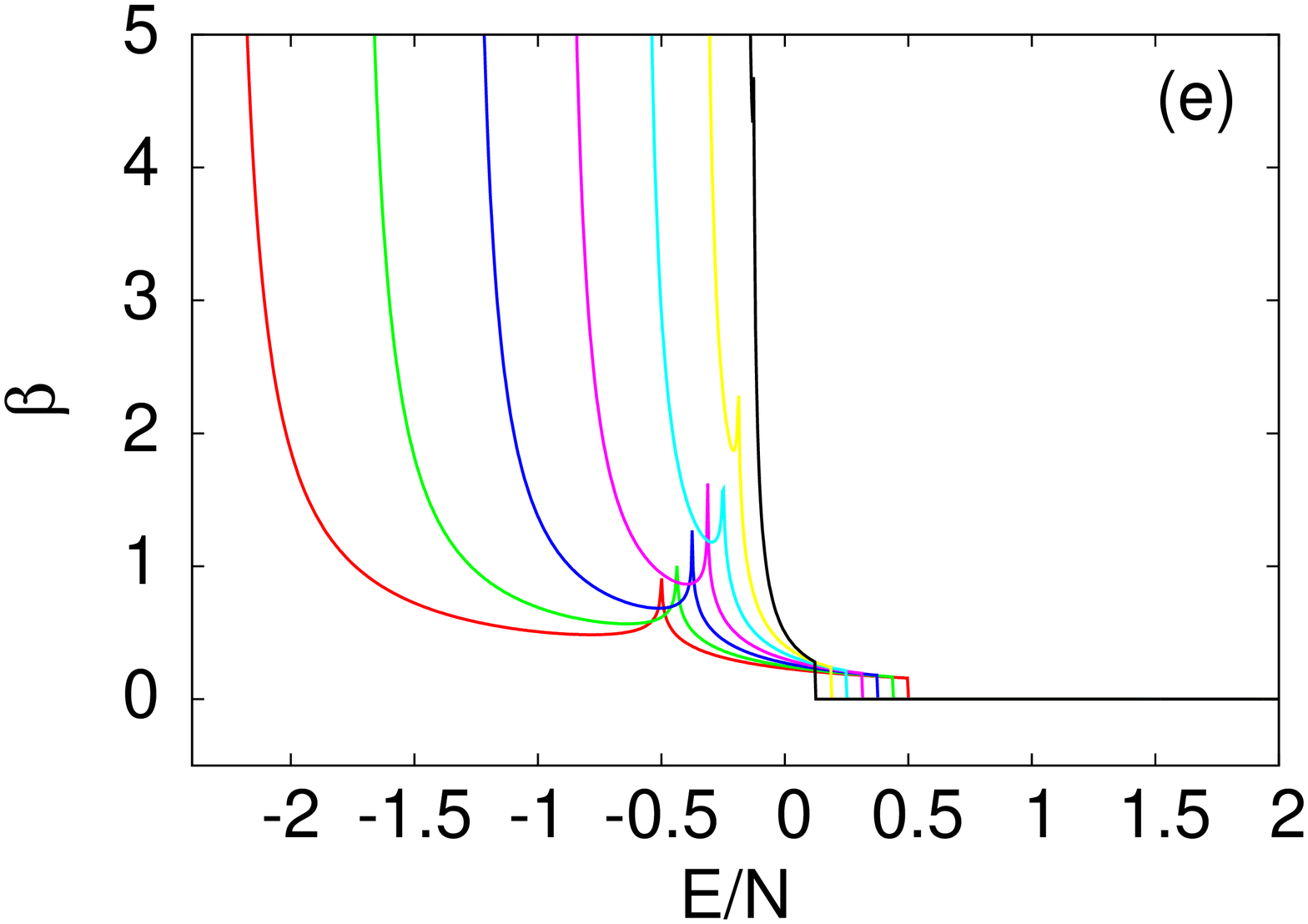} &
\includegraphics[scale=0.3,angle=-90]{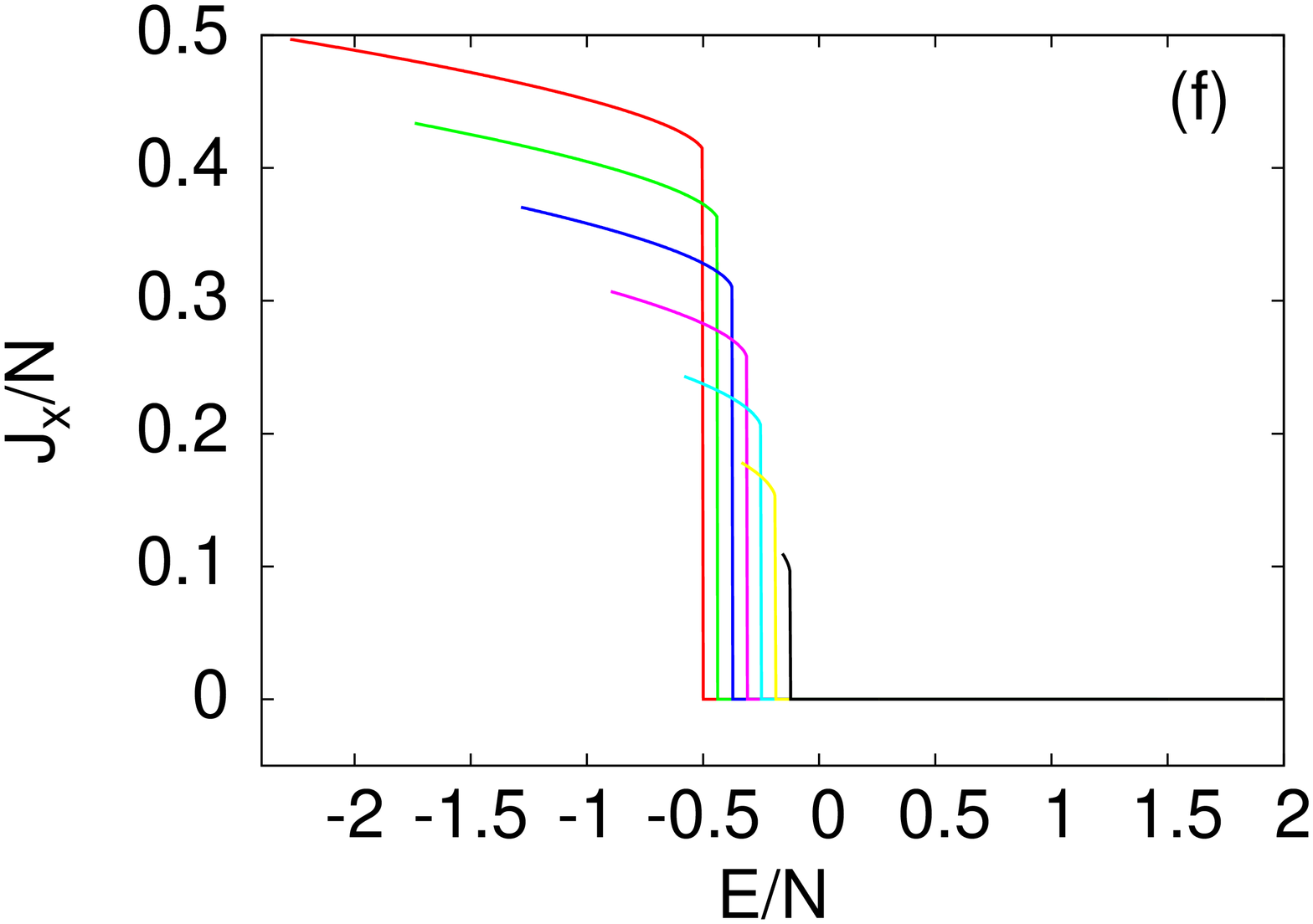} \\
\end{tabular}
\caption{(Color online) Microcanonical calculation for fixed and different values of
  $j$, for $N=10^5$. Panel (a), density of states, $\rho(E)$; panel
  (b), derivative of the density of states $\rho'(E)$; panel (c),
  third component of the angular momentum, $\average{J_z}$; panel (d),
  derivative of the third component of the angular momentum
  $\average{dJ_z/dE}$; panel (e), temperature, $\beta$; panel (f),
  first component of the angular momentum, $\average{J_x}$.  Different
  colors show different values of $j$, $j=2N/16, 3N/16, \ldots,
  8N/16$.}
\label{fig:diff_j}
\end{figure}

Prior to study the ESQPT and the thermal phase transition, we give a
glimpse about the behavior of the different $j$-sectors. In
Fig. \ref{fig:diff_j} we plot the results for the sectors $j=2N/16,
3N/16, \ldots, 8N/16$, with $\omega=\omega_0=1$, $\lambda=1.5$ and
$N=10^5$. In particular, we deal with six different magnitudes: the
density of states, $\rho(E)$; the derivative of the density of states
$\rho'(E)$; the third component of the angular momentum,
$\average{J_z}$; the derivative of the third component of the angular
momentum $\average{dJ_z/dE}$; the temperature, $\beta$; and the first
component of the angular momentum, $\average{J_x}$. All this
magnitudes are calculated by means of the microcanonical formalism;
$\beta$ is the microcanonical temperature
\begin{equation}
\beta = \frac{\partial \log \rho(E)}{\partial E}.
\end{equation}
We can see that the ESQPT occurs at a different energy for each
different $j$-sectors. This is clearly seen in panels (b), (d), (e)
and (f). The first three cases show logarithmic singularities
associated with the derivatives of the density of states and the third
component of the angular momentum \cite{Brandes:13}. It is worth to
mention that this singularity is also present in the microcanonical
temperature $\beta$. Also, note that $\beta$ is not a monotonous
function of the energy; this is a clear signature of the anomalous
thermodynamic behavior of each $j$-sector. Panel (f) shows the
finite jump of the first component of the angular momentum, provided
that the initial state has the parity symmetry broken
\cite{Puebla:13}.

All these facts give important hints to understand the behavior of the
full Hamiltonian, including all the $j$-sectors. If the system remains
thermally isolated and follows a non-trivial time evolution, for
example resulting from a time-dependent protocol $\lambda(t)$, both
the total angular momentum, $J^2$, and the parity, $\Pi$, are
conserved. This entails that the evolution of every $j-\Pi$ sector is
totally independent from the others. The main consequences of this
fact are the following: {\em i)} every $j$-sector is affected by its
ESQPT, showing the dynamical consequences reported in
\cite{Dicke, Puebla:13}; {\em ii)} the behavior of the
total system is the sum of all the sector, weighted by the
corresponding degeneracies $g(N,x)$. In the next section we study the
link between all these features and the thermal phase transition, well
known since more than $40$ years ago \cite{Comer-Duncan:74}.

\subsection{Numerical results: ESQPT versus thermal phase transition}

In order to compare the physics of the isolated Dicke model (for which
$J^2$ and $\Pi$ are conserved quantities) and the Dicke model in
contact with a thermal bath (for which $J^2$ and $\Pi$ are not
conserved), we proceed as follows. On the one hand, we obtain the
microcanonical results, depending on the energy $E$, following the
same procedure than in previous section.  On the other, the 
canonical
calculation depends on $\beta$, and the energy is derived from
Eq. (\ref{eq:energia}). It predicts a critical temperature, given by
Eq. (\ref{eq:critico}), and hence we can obtain the corresponding
values for the critical energy,
\begin{equation}
\average{E_c} = - \left. \frac{\partial \log Z}{\partial \beta} \right|_{\beta_c},
\end{equation}
the critical value of $J_z$
\begin{equation}
\average{J_{z,c}} = \frac{1}{\beta} \left. \frac{\partial \log Z}{\partial \omega_0} \right|_{\beta_c},
\end{equation}
and the derivative of $J_z$
\begin{equation}
\average{\frac{d J_{z,c}}{d E}} = \frac{d}{d E}\frac{1}{\beta} \left. \frac{\partial \log Z}{\partial \omega_0} \right|_{\beta_c} = \frac{\partial}{\partial \beta}\frac{1}{\beta} \left. \frac{\partial \log Z}{\partial \omega_0} \frac{\partial \beta}{\partial E} \right|_{\beta_c}. 
\end{equation}
With the values of the external parameters used in this work,
$\omega=\omega_0=1$ and $\lambda=1.5$, we obtain
\begin{eqnarray}
\beta_c &=& 0.223144, \\
\average{E_c}/N &=& -0.055, \\
\average{J_{z,c}} &=& -0.055 = \average{E_c}/N.
\end{eqnarray}
The derivative of $J_z$ is not defined at the critical temperature
$\beta_c$; it jumps from $0$ to $1$.

\begin{figure}[h]
\includegraphics[scale=0.4,angle=-90]{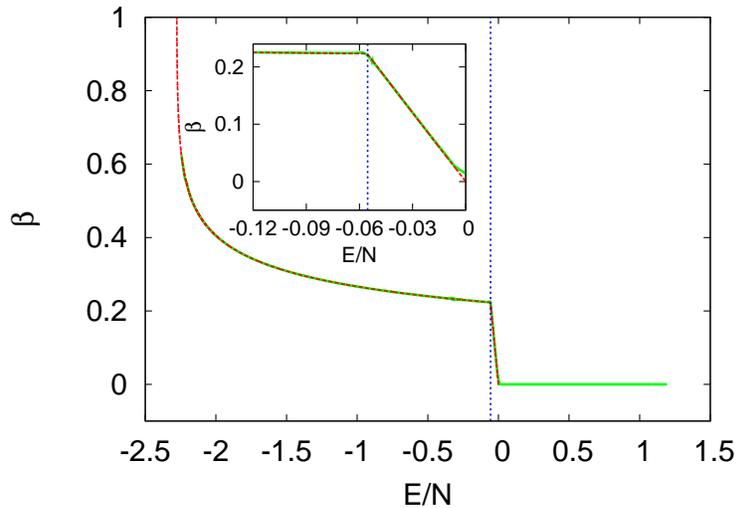}
\caption{(Color online) Temperature $\beta$ vs. energy $\average{E}/N$ obtained by
  means of the microcanonical (solid green line) and the canonical
  (dahsed red line) ensemble. The vertical dashed line shows the
  critical energy $\average{E_c}/N$. The inset shows the same results
  around this critical value.}
\label{fig:beta}
\end{figure}

In Fig. \ref{fig:beta} we plot the temperature $\beta$ in terms of the
energy $\average{E}/N$. We display the microcanonical result by means
of a solid (green online) line, and the canonical result by means of a
dashed (red online) line. The critical value for the energy is shown
by a vertical dashed (blue online) line, and the inset shows a zoom
around the critical energy. Microcanonical calculation is done with
$N=10^5$ particles. The canonical calculation is performed in the
thermodynamical limit, by means of Laplace's method. The results are
pretty different from the ones obtained with the different $j$
sectors. First, we can see that $\beta$ is a monotonous function of
the energy, as one expects from standard thermodynamics. Second,
microcanonical and canonical ensembles give rise to the same results;
particularly, both display the same critical behavior. However, we can
also see an important difference. When the system is put in contact
with a thermal bath, the region with $\average{E}/N > 0$ is
unreachable. In the canonical formalism, the limit $T \rightarrow
\infty$ ($\beta \rightarrow 0$) corresponds with $\average{E}/N
\rightarrow 0$. Hence, if we heat the system by means an external
source of heat, we are restricted to the region with $\average{E}/N <
0$. On the contrary, if the system remains isolated from any
environment, and we {\em heat} the system by means of a mechanical
procedure, for example performing fast cycles between $\lambda_i$ and
$\lambda_f$, we can reach any final energy value. Note that
$\average{E}/N=0$ acts like a second critical energy, since the curve
$\beta(E)$ shows a singularity at this point.

Another remarkable fact is that the logarithmic singularities shown in
panel (e) of Fig. \ref{fig:diff_j} are washed out ---despite
results shown in Fig. \ref{fig:beta} consist of collecting all the $j$
sectors shown in panel (e) of Fig. \ref{fig:diff_j}, weighted by
the corresponding degeneracy according to Eq. (\ref{eq:deg}). On the
other hand, the second singular point, taking place at $E_*^j/N=j/N$
in each $j$-sector, still occurs, at $E_*/N=0$.

\begin{figure}[h]
\includegraphics[scale=0.4,angle=-90]{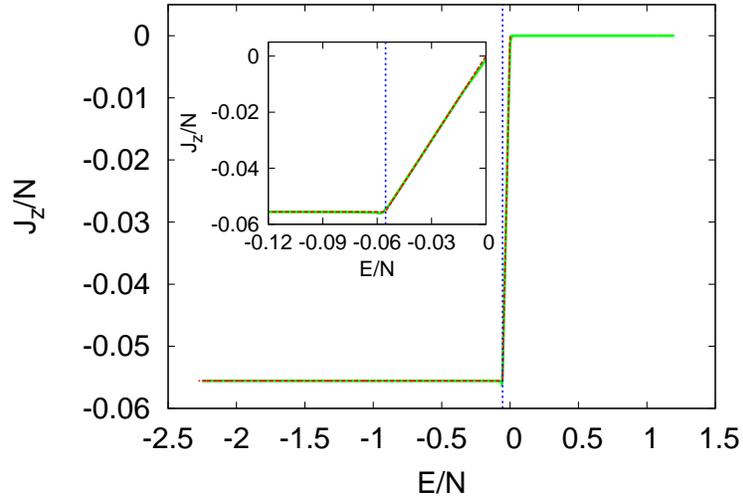}
\caption{(Color online) $J_z$ vs. energy $\average{E}/N$ obtained by
  means of the microcanonical (solid green line) and the canonical
  (dahsed red line) ensemble. The vertical dashed line shows the
  critical energy $\average{E_c}/N$. The inset shows the same results
  around this critical value.}
\label{fig:jz}
\end{figure}

Results for the third component of the angular momentum, $J_z/N$, are
shown in Fig. \ref{fig:jz}. We can see the same kind on
non-analiticity at the critical energy $E_c/N \sim -0.055$ than for
the temperature $\beta$, despite the behavior for each $j$-sector,
shown in panel (c) of Fig. \ref{fig:diff_j}, is totally
different. Furthermore, both microcanonical and canonical calculations
give the same results below $E_*/N=0$. At this value, the
microcanonical ensemble shows a second singular point, and $J_z/N=0$
for $E/N > E_*/N$. It's worth to remark that, despite the consequences
of the ESQPT are not so clear for this magnitude, the minimum
appearing in each $j$-sector just above the critical energy $E_c^j/N$
is not visible in the figure, giving rise to an approximately flat
region $J_z/N \sim -0.055$ for $E<E_c$. However, a zoom around $E_c$
shows that this minimum still exists for finite systems (see below for
more details).

\begin{figure}[h]
\includegraphics[scale=0.4,angle=-90]{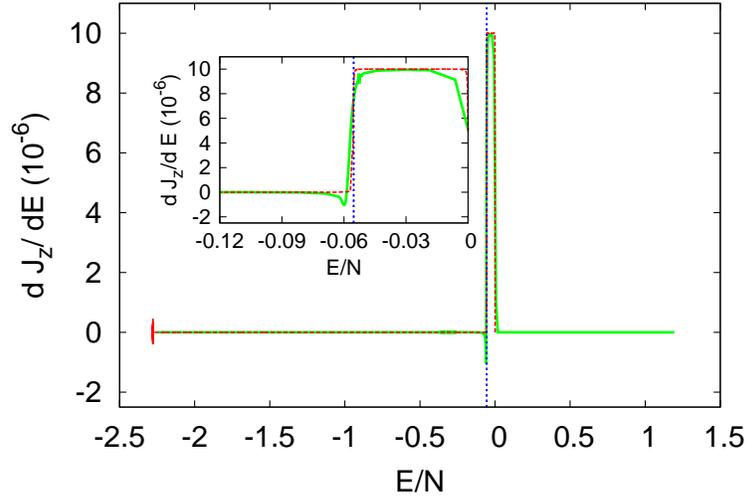}
\caption{(Color online) Derivative of $J_z$ vs. energy $\average{E}/N$ obtained by
  means of the microcanonical (solid green line) and the canonical
  (dahsed red line) ensemble. The vertical dashed line shows the
  critical energy $\average{E_c}/N$. The inset shows the same results
  around this critical value.}
\label{fig:derjz}
\end{figure}

Results for the energy derivative of $J_z$ are shown in
Fig. \ref{fig:derjz}. Again, microcanonical and canonical ensembles
give the same results, below $E_*/N=0$. In this case, we can see a
finite jump at the critical energy $E_c$; the logarithmic
singularities, shown in panel (d) of Fig. \ref{fig:diff_j} are also
ruled out.

\begin{figure}[h]
\includegraphics[scale=0.4,angle=-90]{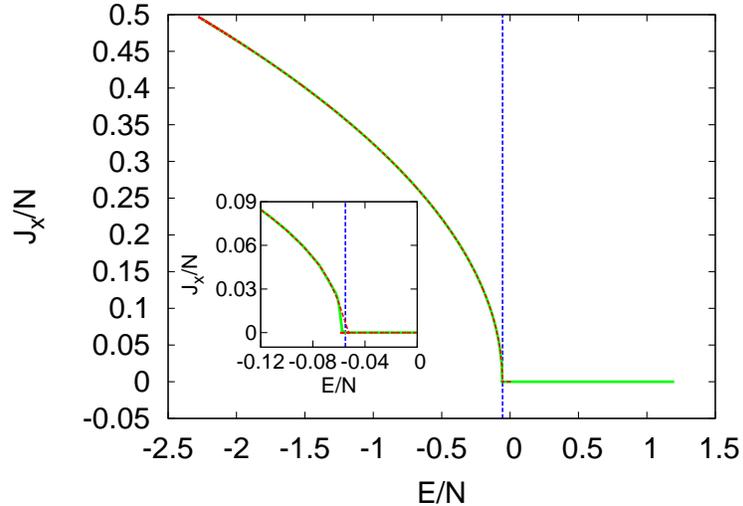}
\caption{(Color online) $J_x$ vs. energy $\average{E}/N$ obtained by
  means of the microcanonical ensemble (solid green line) and the
  canonical ensemble with the symmetry-breaking term $\epsilon J_x$
  (dashed red line). The vertical dashed line shows the critical
  energy $\average{E_c}/N$. The inset shows the same results around
  this critical value.}
\label{fig:jx}
\end{figure}

Finally, results for the first component of the angular momentum,
$J_x/N$ are shown in Fig. \ref{fig:jx}. We depict the
  microcanonical result together with the calculation including the
  symmetry-breaking term, $\epsilon J_x $, described in
  Sec. \ref{sec:symmetry_breaking}. Microcanonical calculations have
been done considering that the parity symmetry is totally broken in
the initial state, and therefore the integrals over the phase space
are restricted to one of the two disjoint regions existing when $E^j/N
< E_c^j/N$ in each $j$-sector. (If we perform the calculations on the
other disjoint region, we obtain the same curve, but with negative
values for $J_x/N$). This observable shows a behavior that is
qualitatively different than the previous ones. The main signature of the ESQPT is still present, but with a different qualitative behavior. $J_x$ is still an order parameter: it changes from $J_x \neq 0$ for $E<E_c$, to $J_x=0$ for $E>E_c$. The main feature of the full Dicke model is that this change in continuous, despite it is discontinuous in every $j$-sector experimenting the ESQPT. 

From all these results, we infer the following conclusions:

\begin{enumerate}

\item Microcanonical and canonical ensembles are equivalent, below the
  singular point located at $E_*/N=0$. This energy constitutes an
  unreachable limit if the system is put in contact with a thermal bath.
  It corresponds to $\beta \rightarrow 0$ (or $T \rightarrow
  \infty$). On the contrary, there is no such a limit if the system
  remains isolated.

\item The main signatures of the ESQPT are ruled out when we collect
  all the $j$-sectors: the logarithmic singularities in the
  derivatives of $\rho$ and $J_z$ are present when the system is
  neither isolated (microcanonical calculation) nor in contact with a
  thermal bath (canonical calculation). As these singularities are
  linked to stationary points in the corresponding semiclassical phase
  space, we can conclude that the relevance of such classical
  structures vanish when all the $j$-sectors are taken into account. A
  possible explanation, compatible with Ref. \cite{Stransky:17}, is that, in this case, the number of effective
  degrees of freedom become infinite, since we have an infinite number
  of $j$-sectors (each one with $f=2$ degrees of freedom) in the
  thermodynamical limit.

\item Contrary to what happens with the other main signatures of the
  ESQPT, the breakdown of the $Z_2$ parity symmetry below the critical
  energy (or temperature), survives. If the system remains isolated
  from any environment, the system behaves as follows. Below the
  critical energy, $E < E_c$, the parity symmetry remains broken if it
  is broken in the initial condition; on the contrary, time evolution
  above the critical energy $E>E_c$ restores the symmetry
  \cite{Puebla:13, Puebla:15}. This entails that the expected value
  $\left< J_x \right>$ keeps relevant information about the initial
  state. On the other hand, parity symmetry becomes spontaneously
  broken if the system is in contact with a thermal bath, as it is
  discussed in Sec. \ref{sec:symmetry_breaking}. The most
  significative result shown in Fig. \ref{fig:jx} is that this
  breakdown exactly coincides with the microcanonical result, when the
  integration over the phase space is restricted to one of the two
  disjoint regions existing for $E<E_c$. That is, thermal fluctuations
  make the system spontaneously {\em choose} one of these to
  possibilities. Hence, it is very worth to note the similarity in the
  behaviour of $\average{J_x}$ in both the excited-state and the
  thermal quantum phase transitions, though the behavior of the system
  is not the same in isolation than in contact with a thermal bath.

\end{enumerate}

\subsection{Results: finite size scaling}

Numerical results in the previous section have been
  obtained following different strategies. When the system is in
  contact with a thermal bath, that is, when we work in the canonical
  ensemble, we make the calculations in the thermodynamical limit,
  relying on the Laplace's method to evaluate the partition
  function. On the contrary, this limit is not explicitely done when
  the system is in isolation and microcanonical ensemble is
  considered. Furthermore, our method is applicable to finite-size
  systems, at least if they are large enough to apply the
  semiclassical approximation to each $j$-sector, and to consider that
  $x=j/N$ is very approximately a continuous variable $x \in
  [0,1/2]$. The aim of this section is to test the applicability of
  our results to systems small enough to be exactly solved by numerical
  diagonalization, and to profit from the analytical results to study
  the finite-size scaling of the critical behavior.

\begin{figure}[h]
\includegraphics[scale=0.4,angle=-90]{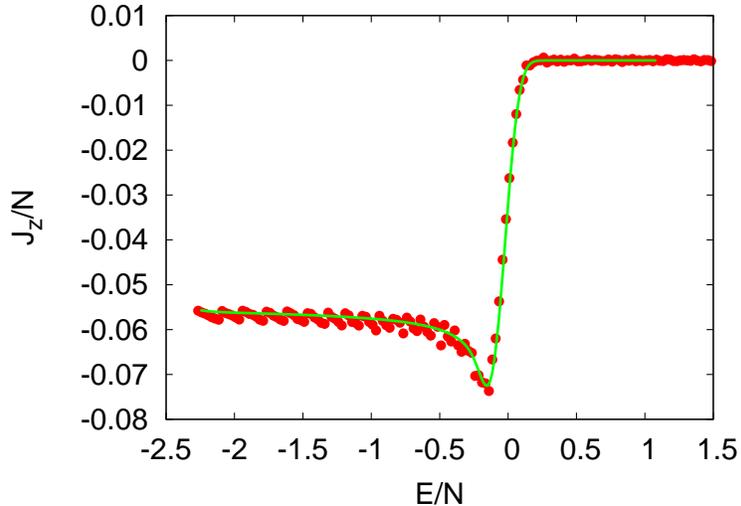}
\caption{(Color online) Exact numerical results (filled circles, red
  online) and microcanonical calculation (solid curve, green online)
  for the expected value of $J_z/N$, in a system with $N=50$ atoms.}
\label{fig:jz_num}
\end{figure}

\begin{figure}[h]
\includegraphics[scale=0.4,angle=-90]{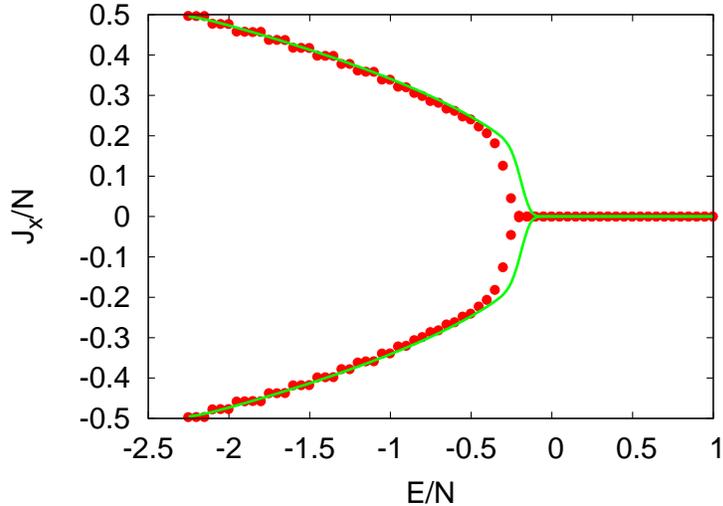}
\caption{(Color online) Exact numerical results (filled circles, red
  online) and microcanonical calculation (solid curves, green online)
  for the expected value of $J_x/N$, in a system with $N=50$
  atoms. Numerical results have been obtained after introducing a
  small symmetry-breaking term, $\epsilon J_x$, with
  $\epsilon=10^{-6}$. For the analytical result, the symmetry-breaking
  term is not introduced, and the two disjoint regions of the phase
  space below the critical energy are integrated separately, to obtain
  the two branches of the theoretical curve.}
\label{fig:jx_num}
\end{figure}

In Fig. \ref{fig:jz_num} we plot the numerical results for
$\left<J_z\right>$ obtained with a system with $N=50$ atoms including
all the $j$-sectors, together with the microcanonical prediction given
by Eq. (\ref{eq:micro_jz}). Numerical results have been obtained as
follows. The Hamiltonian of each $j$-sector, $H_j$, is independently
diagonalized. Then, the expected value in each eigenstate, $J_z (n,j)
= \left< E_n^j \right| J_z \left| E_n^j \right>$ is
calculated. Finally, results for all the $j$-sectors are collected in
an histogram with bins of size $\Delta E/N=0.05$, after considering
the degeneracy of each sector, $g(N,j)$. As the actual number of
photons is unbounded, the dimension of the Hilbert space is infinite,
and hence the diagonalization procedure requires a truncation in the
photonic Hilbert space. For all the calculations shown in this
section, we have taken $n_{\text{max}}=500$ photons, a number large
enough to assure convergence in our results.

The match between theory and numerics is remarkable, taking into
account all the approximations required to obtain the microcanonical
result. At low energies, we see a kind of saw-tooth structure in the
numerical results, which is a consequence of the integrable nature of
the low-lying spectrum of the Dicke model \cite{Relano:16}. Besides
this fact, the microcanonical results give a perfect description of
the model. It is worth to remark the presence of a small dip close to
the critical energy of the ESQPT. As it is discussed below, this dip
is a remanent of the ESQPT and vanishes in the thermodynamical limit.

In Fig. \ref{fig:jx_num} we plot the results for $J_x$, obtained by
means a procedure similar to the previous one. In this case, a small
symmetry-breaking term, $\epsilon J_x $, with
$\epsilon=10^{-6}$, has been introduced for the numerical
diagonalization. As a consequence, the (almost) exact degeneracy of
energy levels below $E_c$ is broken; in this phase, the spectrum
consists of doublets, one level with $\left<J_x\right> >0$, and
another with $\left< J_x \right> < 0$, both with the parity symmetry
totally broken, $\left< \Pi \right> =0$. Hence, to collect the results
for all the $j$-sectors, we have done two different histograms, one
including all the levels with $\left<J_x\right> >0$ and the other
including the levels with $\left<J_x\right> <0$. Also, two
microcanonical integrals, Eq. (\ref{eq:micro_jx}), are performed, each
one restricted to the corresponding disjoint region of the energy
surface. It is worth to remark that the microcanonical integrals have been
performed {\em whithout} including the symmetry-breaking term. Above
the critical energy, only one integration region is considered, since
the energy surface is not splitted anymore. In this region, the
numerical calculations show that every eigenstate has well defined
parity, despite the small symmetry-breaking term introduced in the
Hamiltonian, and that the expected value of $J_x$ is always zero. All
these facts are visible in Fig. \ref{fig:jx_num}. The match between
numerical and microcanonical results is very good, except in the very
surroundings of the critical energy. As large finite-size effects for
this observable have been observed in the highly-symmetric sector
\cite{Puebla:13}, the small discrepancies observed in the figure are
not surprising.

\begin{figure}[h!]
\begin{tabular}{cc}
\includegraphics[scale=0.3,angle=-90]{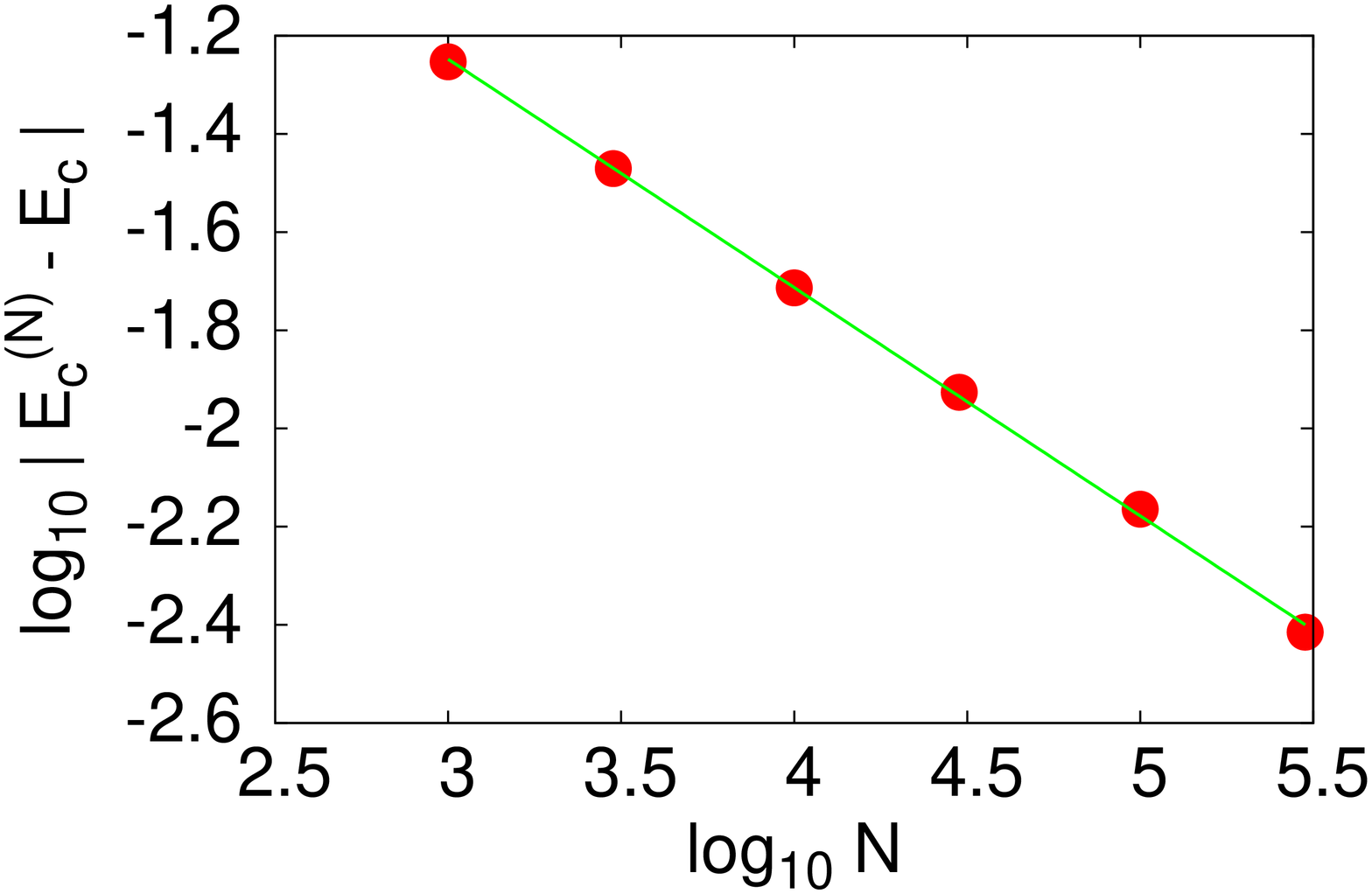} &
\includegraphics[scale=0.3,angle=-90]{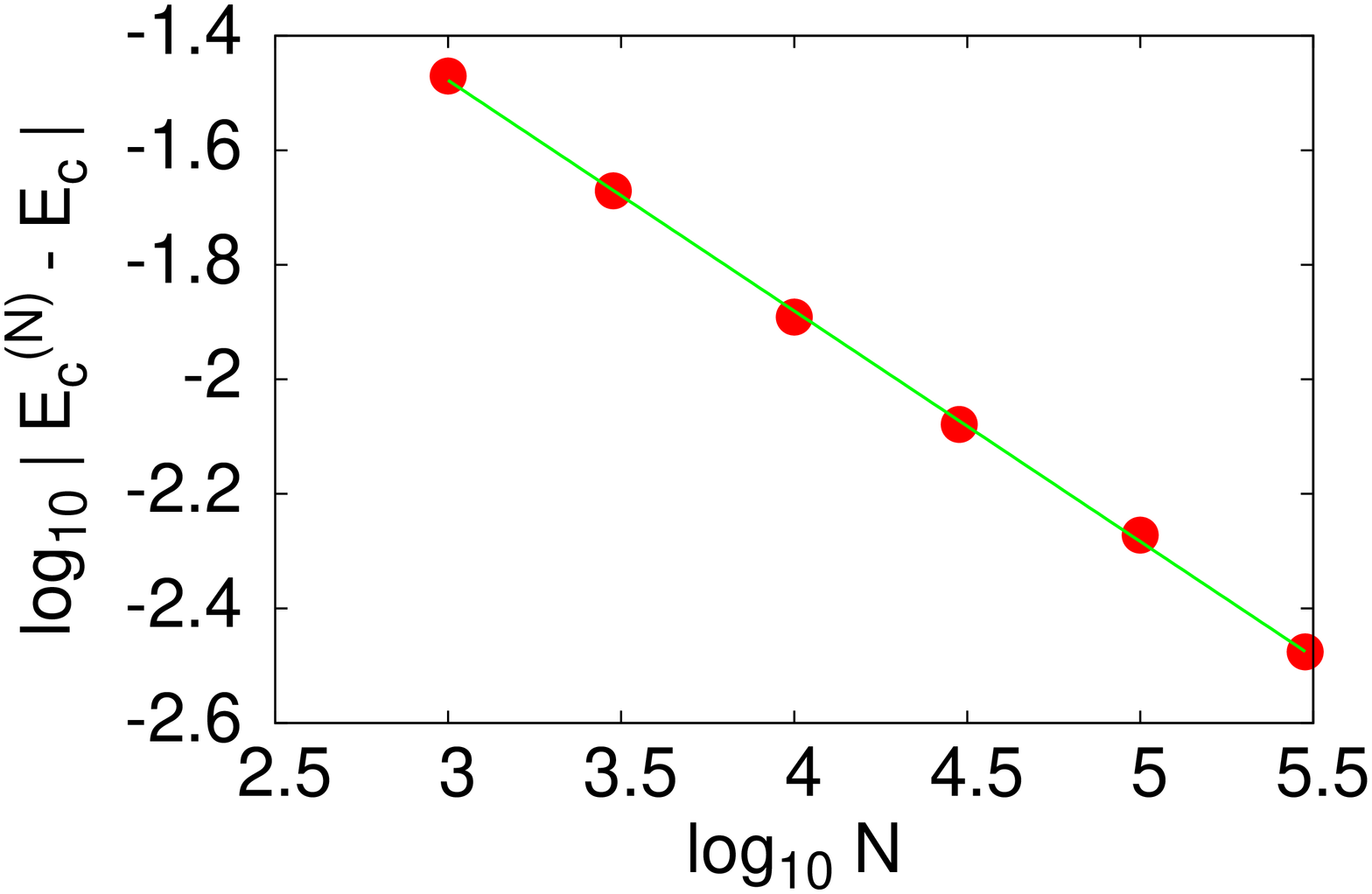} \\
\end{tabular}
\caption{(Color online) Finite size scaling for the critical energy, obtained with $J_z$ (left panel) and $J_x$. Both cases are depicted in a double logarithmic scale. The solid lines represent the least-square fits to straight lines, showing a power-law scaling with the system size.}
\label{fig:scaling_energy}
\end{figure}

We can profit from the previous results to perform a finite-size
scaling analysis of the transition. In particular, we rely on the
theoretical expresions for the microcanonical ensemble to study how
the statistical results depend on the system size $N$. Results for the finite-size precursor of the critical energy
$E_c^{(N)}$ are shown in Fig. \ref{fig:scaling_energy}. We plot the
difference between this precursor and the critical energy obtained by
means the canonical calculation, $E_c^{(N)}- E_c$ versus the size of
the system, in a double logarithmic scale. We also show a straight
line representing the power-law behavior $E_c^{(N)}- E_c \propto
N^{-\alpha}$, with $\alpha(J_z) \sim 0.47$, and $\alpha(J_x) \sim
0.41$. Calculations have been performed as follows. In the left
panel, $E_c^{(N)}$ is estimated as the energy corresponding to the minimum of
$J_z/N$. Though not explicitely shown, this minimum becomes less
pronounced as the system-size grows, vanishing in the thermodynamical
limit. In the right panel, $E_c^{(N)}$ is identified as the energy at
which $J_x/N$ becomes less than $0.01$. This bound is arbitrary, but
we are not interested in quantitative results for each system size
$N$, but in their scaling with the system size. From the results shown
in Fig. \ref{fig:scaling_energy}, we can conclude that the finite size
precursor $E_c^{(N)}$ tends to the critical energy $E_c$, with a
power-law finite-size scaling.

\begin{figure}[h!]
\includegraphics[scale=0.3,angle=-90]{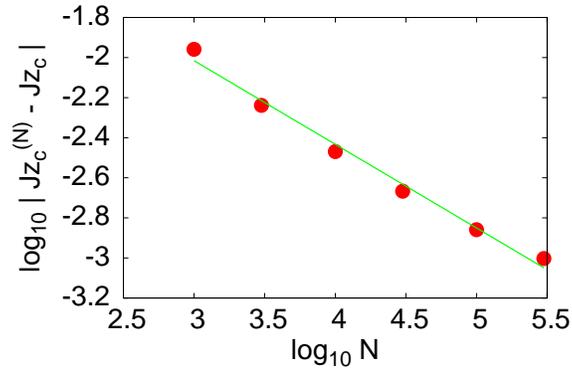} \\
\caption{(color online) Finite size scaling for the critical energy value of $J_z$,
  in a double logarithmic scale. The solid lines represent the
  least-square fits to straight lines, showing a power-law scaling
  with the system size.}
\label{fig:scaling_jz}
\end{figure}

In Fig. \ref{fig:scaling_jz} we show the same results for the critical
value of the third component of the angular momentum, $J_{z,c}^{(N)}-
J_{z,c}$. Though in this case the scaling is not so clean, we still
can conclude that $J_{z,c}^{(N)}- J_{z,c} \propto N^{-\alpha}$, with
$\alpha \sim 0.40$.

\section{Conclusions}

In this work we have analyzed the relationship between the thermal
phase transition and the ESQPT in the Dicke model. First of all, we
have studied the thermodynamics of the model by means of
microcanonical and canonical ensembles, and we have found that both
approaches are incompatible if we considerer just the highly-symmetric
representation, i. e. $j=N/2$. The reason is that the size of the
Hilbert space grows linearly with the number of atoms, $N$, instead of
exponentially. The main consequence is that extensive thermodynamic
magnitudes, like the entropy $S$ or the Helmholtz potential $F$, do
not scale with the number of particles $N$; thermodynamics is
anomalous and the different ensembles are not
equivalent in the thermodynamic limit, $N \rightarrow \infty$. In
order to get a correct description of the thermodynamics properties it
is necessary to include all the $j$-sectors.

To perform the microcanonical calculation including all the $j$
sectors, we have considered that all them can be adequately described
by means of the semiclassical approximation. As a
  consequence, the results for the complete Hilbert space can be
  written as an integral collecting all the sectors, provided that the
  number of particles is large enough. We have shown than $N=50$ atoms
  are enough to guarantee the goodness of this approximation.

We have shown that each $j$-sector is equivalent to the one with
$j=N/2$, but with a smaller effective coupling strength. The main
consequence is that, despite all of them have an ESQPT if the global
coupling strength $\lambda$ is large enough, for any finite value
$\lambda > \lambda_c$ there are a large number of $j$-sectors which
are in the normal phase. To illustrate this fact, we have computed
different magnitudes for different $j$ values: the density of states
$\rho(E)$, the derivative of the density of states $\rho'(E)$, the
third component of the angular momentum $\average{J_z}$, the
derivative of the third component of the angular momentum
$\average{dJ_z/dE}$, the temperature, $\beta$ and the first component
of the angular momentum, $\average{J_x}$.

We have analyzed the relationship between ESQPT and thermal phase transition when all the $j$-sectors are taken into
account. This fact entails that the main signatures of the ESQPT, in
particular the logarithmic singularities in the derivatives of the density of states and the expected value of $J_z$ are ruled out.
However, $\average{J_x}$ still
changes from a value different from zero below the critical energy or
the critical temperature, to zero above them. In particular, we have shown that the parity symmetry is spontaneously broken in the thermodynamical limit, if the system is in contact with a thermal bath (and thus, the canonical ensemble is used). Results obtained in this way coincide with the microcanonical calculations, if the integration in the phase space is restricted to one of the two disjoint regions existing when $E<E_c$. Parity symmetry becomes spontaneously broken at temperatures (or energies) at which the underlying semiclassical space is splitted in two regions; thermal fluctuations make the system {\em choose} one of the two existing disjoint regions.

Finally, we have also discussed the main physical differences between
the Dicke model in isolation and in contact with a thermal
bath. Despite both the microcanonical and canonical descriptions
mainly coincide in the thermodynamic limit, one important difference
remains. If the system is in contact with a thermal bath, that is, if
it is described by means of the canonical ensemble, the energy
$E_*/N=0$ constitutes an upper bound; this energy implies $T
\rightarrow \infty$, and thus cannot be exceeded in any experiment. On
the contrary, if the system remains thermally isolated and is {\em
  heated} by means of the Joule effect, for example by quenching
$\lambda_i \rightarrow \lambda_f \rightarrow \lambda_i$ repeatedly,
the limit $E_*/N=0$ can be exceeded; in other words, $E/N>0$ are
accesible in the microcanonical description.

\begin{acknowledgments}

  The authors gratefully acknowledge discussions with
  M. A. Bastarrachea-Magnani, S. Lerma-Hern\'andez, J. G. Hirsch and
  P. Cejnar. A. R. is supported by Spanish Grants No. FIS2012-35316
  and FIS2015-63770-P (MINECO/FEDER) and P. P. F. is supported by 
Spanish Grant No. FIS2014-53448-C2-1-P (MINECO/FEDER).

\end{acknowledgments}

\end{document}